\documentclass[prl,unsortedaddress,groupedaddress,twocolumn,amsmath,amsfonts,amssymb,showpacs,floatfix,nofootinbib]{revtex4}
\usepackage{graphicx}

\usepackage{amsmath}
\usepackage{bm}
\usepackage{color}

\graphicspath{%
    {converted_graphics/}
    {/}
}
\begin{document}

\title{Radiative capture  reactions via indirect method}
\author{A. M. Mukhamedzhanov$^{1}$}
\email{akram@comp.tamu.edu}
\author{G. V. Rogachev$^{1,2}$}
\email{rogachev@tamu.edu}
\affiliation{$^{1}$Cyclotron Institute, Texas A$\&$M University, College Station, TX 77843, USA}
\affiliation{$^{2}$Department of Physics and Astronomy, Texas A$\&$M University, College Station, TX 77843, USA}

\date{Today}

\begin{abstract}
Many radiative capture reactions of astrophysical interest occur at such low energies that their direct measurement is hardly possible. In this paper we address the indirect method, which can provide a powerful technique to obtain information about radiative capture reactions at astrophysically relevant energies. The idea of the indirect method is to use the indirect reaction $A(a, s\,\gamma)F$ to obtain information about the radiative capture reaction $A(x,\,\gamma)F$, where $a=(s\,x)$ and $F=(x\,A)$. The main advantage of using the indirect reactions is the absence of the penetrability factor in the channel $x+A$, which suppresses the low-energy cross sections of the  $A(x,\,\gamma)F$ reactions and does not allow to measure these reactions at astrophysical energies. A general formalism to treat indirect resonant radiative capture reactions is developed when only a few intermediate states do contribute and statistical approach cannot be applied. The indirect method requires coincidence measurements of the triple differential cross section, which is a function of the photon scattering angle, energy  and scattering angle of the outgoing spectator-particle $s$. 
Angular dependence of the triple differential cross section at fixed scattering angle of the spectator $s$ is the angular $\gamma-s$ correlation function.
Using indirect resonant radiative capture reactions one can obtain the information about important astrophysical resonant radiative capture reactions, like $(p,\,\gamma), \,\,(\alpha,\,\gamma)$ and $(n,\,\gamma)$ on stable and unstable isotopes. The indirect technique makes accessible low-lying resonances, which are close to the threshold, and even subthreshold bound states located at negative energies. 
In this paper, after developing the general formalism, we demonstrated the application of the  indirect reaction ${}^{12}{\rm C}({}^{6}{\rm Li},d\,\gamma){}^{16}{\rm O}$ proceeding through $1^{-}$ and $2^{+}$ subthreshold bound states and resonances to obtain the information about the ${}^{12}{\rm C}(\alpha,\,\gamma){}^{16}{\rm O}$ radiative capture at astrophysically most effective energy $0.3$ MeV what is impossible using standard direct measurements. Feasibility of the suggested approach is dicussed.
\end{abstract}

\maketitle

\section{Introduction}
\begin{figure}[tbp] 
  \centering
 \includegraphics[bb=0 258 794 595,width=4.0in,height=2.41in]{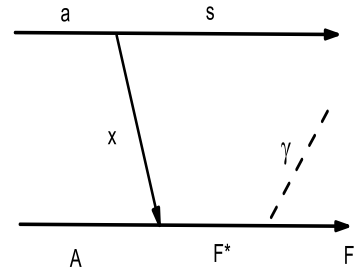}
  \caption{Pole diagram describing the indirect radiative capture reaction proceeding through
the intermediate excited state $F^{*}$. }
  \label{fig:fig_polediagram1}
\end{figure}

Two-step transfer reactions $A(a,s\,\gamma)F$, proceeding through the intermediate subthreshold bound states or resonances $F^{*}= (x\,A)^{*}$, with the subsequent decay of the excited state $F^{*} \to F + \gamma$, provide a powerful indirect technique to study radiative capture processes $A(s,\,\gamma)F$ and, in particular, the astrophysical radiative capture reactions. The mechanism of such processes
is shown in Fig. \ref{fig:fig_polediagram1}, where $a=(s\,x)$ and $F=(x\,A)$ are the ground bound states of $a$ and $F$.
Such indirect reactions allow one to invade into the region previously unthinkable if we would rely only on direct measurements. Among of the important reactions, which require a broader approach than only direct measurements, are low-energy astrophysical radiative capture processes, such as $(p,\,\gamma), \,\,(\alpha,\,\gamma)$ and $(n,\,\gamma)$ on stable and unstable isotopes performed in direct and inverse kinematics. Among these reactions, without any doubt, is the most important one, the so-called, "holy grail" reaction ${}^{12}{\rm C} + \alpha \to  {}^{16}{\rm O}(0^{+}, E_{x}=0.0 {\rm MeV}) + \gamma$, which dominates the helium burning in red giants \cite{rolfssydney}. The indirect reactions provide a perfect tool to study radiative capture reactions at astrophysically relevant energies.

In this paper we present the theory of the indirect method to treat the resonant radiative capture processes when only a few subthreshold bound states and resonances are involved
and statistical methods cannot be applied. The developed formalism is based on the generalized multi-level  $R$-matrix approach and surface integral formulation of the transfer reactions, which are the first stage of the indirect reaction mechanism described by  the diagram in Fig. 
\ref{fig:fig_polediagram1} \cite{reviewpaper}. We describe also the technology of the indirect radiative capture experiment.
There are many papers devoted to the angular correlation of the photons emitted in the nuclear transfer reactions
with the final nuclei, see, for example, \cite{Austern} and references therein. Here we apply the generalized $R$-matrix to develop the formalism allowing one to study the photon's angular distribution correlated with the scattering angle of one of the final nuclei formed in the transfer reaction.

Developing quite a general formalism we keep in mind, in particular, the application of the method for the ${}^{12}{\rm C}({}^{6}{\rm Li},d\,\gamma){}^{16}{\rm O}$ reaction, which can provide important information about the astrophysical 
$\,{}^{12}{\rm C}(\alpha,\,\gamma){}^{16}{\rm O}$ process. Because this astrophysical reaction is contributed by two interfering subthreshold resonances and resonances at positive energies, our developed formalism of the indirect processes include subthreshold and real resonances.
A subthreshold bound state (a bound state, which  is close to the threshold) reveals itself as a subthreshold resonance at low-energy  reactions. Subthreshold resonances play an important role in many astrophysical processes. Often, using direct measurements  it is quite difficult or impossible to reach the astrophysically relevant energy region where the subthreshold resonances manifest themselves. However,  the region, where the contribution of the subthreshold resonances is important, can be reached using indirect reactions  \cite{reviewpaper}.

Numerous attempts to obtain the astrophysical factor of the $\,{}^{12}{\rm C}(\alpha,\,\gamma){}^{16}{\rm O}$ reaction, both experimental and theoretical, have been made for almost 50 years \cite{redder,barker78,barker87,barkerkajino,azuma,brune1999,kunz,assuncao,Schuermann,gai2015,sayre,gai2013,brune2013,gai2014,avila}. This reaction is contributed by interfering $E1$ and $E2$ transitions. 
The $E1$ transition is complicated by the interference of the capture through the wing of the subthreshold $1^{-}$ resonance at $-0.045$ MeV with the low-energy tail  of the resonance  $1^{-},\,E_{\alpha\,{}^{12}{\rm C}}=2.423$ MeV, where $E_{\alpha\,{}^{12}{\rm C}}$ is the $\alpha-{}^{12}{\rm C}$ relative kinetic energy. The $E2$ transition is dominated by the capture to the ground state of ${}^{16}{\rm O}$ through the wing of the subthreshold  bound state $2^{+},E_{\alpha\,{}^{12}{\rm C}}=-0.245$ MeV.  In addition, to fit the experimental data, usually
a few artificial levels are added to fit $E1$ and $E2$ data \cite{barkerkajino,brune1999}. The difficulty of the direct measurements of the $E1$ transition can be easily understood if even in the peak of the resonance at
$1^{-},\,E_{\alpha\,{}^{12}{\rm C}}=2.423$ MeV the cross section is only about $40-50$ nb. Moreover , the $E1$ transition from $1^{-}$ states to the ground state of ${}^{16}{\rm O}$ is isospin forbidden for $T=0$ components and is possible only due to the small admixture of the $T=1$ components.

Extremely small penetrability factor at  $E_{\alpha\,{}^{12}{\rm C}} \leq 1$ MeV makes it impossible or very difficult to measure the astrophysical factor for the $\,{}^{12}{\rm C}(\alpha,\,\gamma){}^{16}{\rm O}$ reaction at energies $E_{\alpha\,{}^{12}{\rm C}} \leq  1$ MeV  with reasonable accuracy. For the sensitivity of the extracted  astrophysical factor from the existing data see  works \cite{gai2013,gai2014,gai2015}. Note that from the astrophysical point of view  the required uncertainty  of this astrophysical factor at $E_{\alpha\,{}^{12}{\rm C}} \sim 0.3$ MeV should be $\leq 10\%$. New gamma-ray facilities in the USA and Romania are supposed to measure the astrophysical factor for the $\,{}^{12}{\rm C}(\alpha,\,\gamma){}^{16}{\rm O}$ reaction down to $1$ MeV.

In this paper we discuss a completely new method of measuring the astrophysical factor $S(E_{\alpha\,{}^{12}{\rm C}})$ for the $\,{}^{12}{\rm C}(\alpha,\,\gamma){}^{16}{\rm O}$ reaction down to astrophysical energies $\sim 300$ keV. This method is based on the coincidence measurements of the deuterons and the photons from the indirect reaction ${}^{12}{\rm C}({}^{6}{\rm Li},d\,\gamma){}^{16}{\rm O}$. In the indirect method the absolute value of the triple differential cross section is determined by its normalization to the available direct data at higher energies.

The suggested technique allows one not only to determine the astrophysical $S$ factor down to energies $E_{\alpha\,{}^{12}{\rm C}}  \sim 0.3$ MeV  but also the interference pattern between the subthreshold bound state and higher resonance for the $E1$ transition. Another advantage of the method is that the problem of the $R$-matrix relative phase shift of $E1$ and $E2$ transitions do not appear in the indirect method.
The method, which we address here, can be used for a broader type of radiative capture experiments  $A(a,\,s\,\gamma)F$ proceeding through the subthreshold and real resonances.

\section{Theory}

To measure the cross section of the binary process
\begin{equation}
x + A \to F^{*} \to \gamma +F
\label{reaction1}
\end{equation}
proceeding through the intermediate resonance $F^{*}$  at astrophysical energies we suggest to measure the  reaction (two-body to three-body process ($2 \to 3$ particles)):
\begin{align}
a + A \to s + F^{*} \to s + \gamma + F
\label{THMreaction1}
\end{align}
in the vicinity of the quasi-free (QF) kinematics.
Here the incident particle, $a = (s\,x)$, which has a dominant cluster structure, is accelerated at energies above the Coulomb barrier. The reaction (\ref{THMreaction1}) is a two-stage process.
On the first stage the transfer reaction $a + A \to s+ F^{*}$ populating the wing of the subthreshold bound state at $E_{xA} >0$ or the real resonance  occurs. On the second stage the excited state $F^{*}$ decays to the ground state $F\,$ by emitting a photon. From the measured energy dependence of the cross section of the reaction  (\ref{THMreaction1}), the energy dependence of the binary sub-process (\ref{reaction1}) is determined.  By normalizing the measured cross section to the available direct one(s) measured at higher energies with better accuracy one can get the absolute value of the astrophysical $S$ factor at low energies.

The mechanism of the indirect reaction  shown schematically in Fig. (\ref{fig:fig_polediagram1})  gives the dominant contribution to the cross section in a restricted region of the three-body phase space when the relative momentum of the fragments $s$ and $x$ is zero (the QF kinematical condition) or less than the wave number of the bound state $\,a=(s\,x)$. Since the transferred particle $x$  is virtual, its energy and momentum are not related by the on-shell equation, that is, $\,E_{x}\not= k_{x}^{2}/(2\,m_{x})$.

The main advantage of the indirect method is that the penetrability factor in the entry channel of the binary reaction (\ref{reaction1}) is not present in the expression for the indirect reaction cross section. It allows one to measure the resonant reaction  (\ref{reaction1}) cross section at astrophysically relevant energies at which direct  measurements are impossible or extremely difficult  because of the presence of the penetrability factor in the binary reaction cross section. Moreover, the indirect method allows one to measure the cross section of the binary reaction (\ref{reaction1}) even at negative $E_{xA}$ owing to the off-shell character of the transferred particle $x$ in reaction (\ref{THMreaction1}).

The expression for the amplitude of the transfer reaction (\ref{THMreaction1}) ( for $x=n$)  in the surface integral approach and distorted wave Born approximation (DWBA) was derived in \cite{muk2011}. It is assumed, similar to the THM \cite{reviewpaper}, that only the energy dependence  of the cross section of the reaction (\ref{THMreaction1}) is  measured, while its absolute value  is determined by normalizing  the  cross section of the reaction (\ref{THMreaction1})  to the available direct experimental data at higher energies. That is why it makes sense to use the plane wave approximation to get the indirect reaction amplitude.
In this paper we, for the first time, present the general equations of the indirect reaction triple differential and double cross sections to be used for the analysis of the radiative reactions proceeding through the subthreshold and isolated resonances.
The system of units in which $\hbar=c=1$ is being used throughout of the paper.

\subsection{Indirect reaction amplitude for the  resonant radiative capture}

Let us consider the radiative capture reaction  (\ref{reaction1}) proceeding through the wing (at $E_{xA} >0$) of the subthreshold bound state (aka subthreshold resonance) $F^{*} = F^{(s)}$, where  $F^{(s)} =(x\,A)^{(s)}$ or real resonance at $E_{xA} >0$. We assume that both can decay to the ground state $F=(xA)$. To measure the cross section of this reaction at astrophysically relevant energies where subthreshold resonances can be important, for the reasons explained above,  we  use the indirect reaction (\ref{THMreaction1}).
First, we derive the reaction amplitude of the indirect radiative capture process and then the triple differential cross section of reaction (\ref{THMreaction1}). After that, by integrating over the angles of the emitted photons, we get the double differential cross section.
The interference of  the subthreshold bound state and the resonance, which both decay to the ground state
$F=(xA)$, is taken into account. Evidently that this case can be applied for the $E1$ and $E2$ transitions of the reaction ${}^{12}{\rm C}(\alpha,\gamma){}^{16}{\rm O}$.
To describe the radiative capture to the ground state through two interfering states  we need to revoke the single channel, two-level generalized $R$-matrix equations developed for the three-body reactions $2\,\,{\it particles} \to 3\,\,{\it particles}$ \cite{reviewpaper,muk2011}.  
We also take into account the interference of transitions with different multipolarities $L$. Thus, we take into account the interference of transitions from different levels with the same multipolarity and interference of transitions from different levels with different multipolarities.  

The indirect reaction described by the diagram of Fig. (\ref{fig:fig_polediagram1})  proceeds as a two stage process. The first part is transfer of particle $x$ (stripping process) to the excited state $F_{\tau},\,\,\tau=1,2$, where $F_{1} =F^{(s)}$ is the subthreshold resonance and $F_{2}$ is the resonance state at $E_{xA} >0$.  No gamma is emitted during the first stage. On the second stage the excited state $\,F_{\tau}$ decays to the ground state $\,F=(xA)$ by emitting a photon.
Then the indirect reaction amplitude followed by the photon emission from the intermediate subtheshold resonance and resonance takes the form (with fixed projections of the spins of the initial and final particles including the photon):
\begin{align}
M_{M_{a}\,M_{A}}^{M_{s}\,M\,M_{F}} = \sum\limits_{\tau, \nu =1}^{2}\,\sum\limits_{M_{F_{\nu}^{(s)}}\,M_{F_{\tau}^{(s)}}}\, V_{M_{F_{\nu}^{(s)}}\,\nu}^{M_{F}\,M}\,{\rm {\bf A}}_{\nu\,\tau}\,M_{M_{a}\,M_{A}\,\tau}^{M_{s}\,M_{  F_{\tau}^{(s)}}}.
\label{THradamplfull1}
\end{align}
Here $M_{i}$ is the projection of the spin $J_{i}$ of the particle $i$,   $M_{F_{\tau}^{(s)}}$ is the projection of the spin $J_{F^{(s)}}$ of the subthreshold resonance ($\tau =1$) and resonance ($\tau=2$), $M$ is the projection of the angular momentum of the emitted photon. Also $2$ is the number of the level included. We assume that the spins of the subthreshold resonance and real resonance are equal $F_{1} = F_{2}= F^{(s)}$ and these resonances do interfere. At the moment we confine ourselves by transition with one multipolarity $L$. That is why the index $L$ is omitted. Later on we take into account transitions with different $L$.
$M_{M_{a}\,M_{A}\,\tau }^{M_{s}\,M_{F_{\tau}^{(s)}}}$ is the amplitude of the direct transfer reaction 
\begin{align}
a+ A \to s+ F_{\tau}
\label{transfreact1}
\end{align} 
populating the intermediate excited state $F_{\tau}$. Reaction (\ref{transfreact1}) is the first stage of the indirect reaction (\ref{THMreaction1}).
$V_{\nu}$ is the amplitude of the radiative decay of the excited state  $\,F_{\nu} \,\,(\nu=1,2)$ to the ground state $\,F=(xA)$, $\,{\rm {\bf A}}_{\lambda\,\tau}\,$ is the matrix element of the level matrix in the $\,R$-matrix method.  

In the prior form of the plane-wave approximation $M_{M_{a}\,M_{A}\,\tau }^{M_{s}\,M_{ F_{\tau}^{(s)}}}$  takes the form:
\begin{align}
M_{M_{a}\,M_{A}\,\tau}^{M_{s}\,M_{F_{\tau}^{(s)}}}({\rm {\bf k}}_{sF_{\tau}},{\rm {\bf k}}_{aA}) = <\chi_{{\rm {\bf k}}_{sF_{\tau}}}^{(0)}\,\Phi_{\tau}\,\big|V_{xA}\big|\varphi_{sx}\,\varphi_{A}\,\chi_{{\rm {\bf k}}_{aA}}^{(0)}>.
\label{Mtau1}
\end{align}
Here, $\Phi_{1}$ is the bound-state wave function of the subthreshold bound state $F_{1}=(xA)^{(s)}$, $\,\Phi_{2}$  is the $F_{2}$ resonance wave function, $\,\varphi_{sx}$ and $\,\varphi_{A}$ are the bound-state wave functions of $a=(s\,x)$ and $A$, correspondingly, $\,\chi_{ {\rm {\bf k}}_{aA} }^{(0) }= e^{ i\,{\rm {\bf k}}_{aA} \cdot {\rm {\bf r}}_{aA} }$ and $\chi_{{\rm {\bf k}}_{sF_{\tau}}}^{(0)} = e^{i\,{\rm {\bf k}}_{sF_{\tau}} \cdot {\rm {\bf r}}_{sF}}$ are the planes waves in the initial and final states of the reaction (\ref{transfreact1}), correspondingly, $\,{\rm {\bf r}}_{ij}$ is the radius-vector connecting  the centers of mass of nuclei $\,i$ and $j$,  $\,\,{\rm {\bf k}}_{aA}$ is  the $a-A$ relative momentum in the initial state of the reaction (\ref{transfreact1})    and  $\,\,{\rm {\bf  k}}_{sF_{\tau}}$ is the $s-F_{\tau}$  relative momentum in the final state of this transfer reaction,  $\,V_{xA}$ is the $\,x-A$ interaction potential.

In the matrix element $M_{M_{a}\,M_{A}\,\tau}^{M_{s}\,M_{F_{\tau}^{(s)}}}({\rm {\bf k}}_{sF_{\tau}},{\rm {\bf k}}_{aA})$ we introduce in the bra state the projection operator $\sum\limits_{n}|\varphi_{A_{n}}><\varphi_{A_{n}}|$, where sum over $n$ is taken over the bound and continuum states of nucleus $A$. In the projection operator we keep only the projection on the ground state of $A$. Then   Eq. (\ref{Mtau1}) can be rewritten as
\begin{align}
M_{M_{a}\,M_{A}\,\tau}^{M_{s}\,M_{F_{\tau}^{(s)}}}({\rm {\bf k}}_{sF_{\tau}},{\rm {\bf k}}_{aA}) = <\chi_{{\rm {\bf k}}_{sF_{\tau}}}^{(0)}\,\Upsilon_{\tau}\,\big|{\overline V}_{xA}\big|\varphi_{sx}\,\chi_{{\rm {\bf k}}_{aA}}^{(0)}>.
\label{Mtau2}
\end{align}
Here,
\begin{align}
&\Upsilon_{\tau}({\rm {\bf r}}_{xA})= <\varphi_{A}\,\big|\Phi_{\tau}> =      \sum\limits_{m_{j_{i}}\,m_{l_{i}} }<j_{i}\,m_{j_{i}}\,\,l_{i}\,m_{l_{i}}\big| J_{F^{(s)}}\,M_{F_{\tau}}^{(s)}>\nonumber\\
& \times<J_{x}\,M_{x}\,\,J_{A}\,M_{A} \big| j_{i}\,m_{j_{i}}>\,\Upsilon_{ \tau\,j_{i}\,l_{i}\,J_{F^{(s)}}}(r_{xA})\,Y_{ l_{i}\,m_{l_{i}} }({\rm {\bf {\hat r}}}_{xA})
\label{Upsilon1}
\end{align}
is the projection of the wave function $\Phi_{\tau}$ on the ground state wave function of $A$,  $\,\Upsilon_{ \tau\,j_{i}\,l_{i}\,J_{F^{(s)}}}(r_{xA})$ is its radial part,  $\,j_{i}\,(m_{j_{i}})$ is the channel spin (its projection) of $x+A$ and $\,l_{i}\,(m_{l_{i}})$ is their orbital angular momentum (its projection) at which the subthreshold  resonance and resonance occur in the channel $x+A$, $<j_{i}\,m_{j_{i}}\,\,l_{i}\,m_{l_{i}}\big| J_{F^{(s)}}\,M_{F_{\tau}}^{(s)}>$ is the Clebsch-Gordan coefficient.

$\Upsilon_{\tau\,j_{i}\,l_{i}\,J_{F^{(s)}}}(r_{xA})$  is the radial part of  $\Upsilon_{\tau}({\rm {\bf r}}_{xA})$  in the state $\tau$ with  the channel spin $j_{i}$ and the orbital angular momentum $l_{i}$.  Since we assume that both levels $\tau=1$ and $\tau=2$ do interfere, $j_{i}$ and $l_{i}$ are the same for both levels.  We assume that only one $j_{i}$ and $l_{i}$ contribute to the reaction.
It is important to underscore that, although the subthreshold resonance is located at $E_{xA}= -\varepsilon_{xA}^{(s)}$, the capture occurs to its wing at $E_{xA} >0$. Hence, $\Upsilon_{ 1\,j_{i}\,l_{i}\,J_{F^{(s)}}}(r_{xA})$  is described by the resonance radial wave function, which we take in the $R$-matrix form.  We also take the radial overlap $\Upsilon_{ 2\,j_{i}\,l_{i}\,J_{F^{(s)}}}(r_{xA})$ in the form of the $R$-matrix resonant wave function. It has been shown in \cite{reviewpaper} that in the surface integral approach the dominant contribution to the prior form of the transfer reaction amplitude comes from the external region $r_{xA}  \geq R_{xA}$. In the external region we take the resonance wave function as
\begin{widetext}
\begin{align}
\Upsilon_{ \tau\,\,j_{i}\,l_{i}\,J_{F^{(s)}}}(r_{xA}) =  
\sqrt{\frac{\mu_{xA}}{k_{xA}}\,\Gamma_{\tau\,j_{i}l_{i}J_{F^{(s)}}}}\,e^{-i\,\delta^{hs}_{l_{i}}}\,\frac{O_{l_{i}}(r_{xA})}{r_{xA}}.
\label{extreswf1}
\end{align}
\end{widetext}
At $r_{xA}=R_{xA}$ we get 
\begin{align}
\Upsilon_{ \tau\,\,j_{i}\,l_{i}\,J_{F^{(s)}}}(R_{xA})=\sqrt{\frac{2\,\mu_{xA}}{R_{xA}}}\,\gamma_{\tau\,j_{i}l_{i}J_{F^{(s)}}}.
\label{UpsilonRxA1}
\end{align}

$\Gamma_{\tau\,j_{i}l_{i}J_{F^{(s)}}}$ is the formal resonance width in the $R$-matrix approach for the level $\tau$, which is related to the reduced width   
amplitude $\gamma_{\tau\,j_{i}l_{i}J_{F^{(s)}}}$ of the level $\tau$ as
\begin{align}
\Gamma_{\tau\,j_{i}l_{i}J_{F^{(s)}}}=2\,P_{l_{i}}(E_{xA},\,R_{xA})\,\gamma_{\tau\,j_{i}l_{i}J_{F^{(s)}}}^{2}.
\label{Gammaredwidth1}
\end{align}
Here,
$P_{l_{i}}(E_{xA},\,R_{xA})$ is the barrier penetrability factor, $R_{xA}$ is the channel radius, $\,O_{l_{i}}(k_{xA},\,r_{xA})=i\, F_{l_{i}}(k_{xA},\,r_{xA}) + G_{l_{i}}(k_{xA},\,r_{xA}) $ is the outgoing spherical wave in the partial wave $l_{i}$, $\,F_{l_i}$  and $\,G_{l_i}$are the Coulomb regular and singular solutions, $\delta_{l_{i}}^{hs}$ is the $R$-matrix hard sphere scattering phase shift.
Equation (\ref{Gammaredwidth1}) holds at $E_{xA} >0$ both for the subthreshold resonance
and resonance. 

The observable resonance width is expressed in terms of the observable reduced width by
\begin{align}
{\tilde \Gamma}_{\tau\,j_{i}l_{i}J_{F^{(s)}}}=2\,P_{l_{i}}(E_{xA},\,R_{xA})\,{\tilde \gamma}_{\tau\,j_{i}l_{i}J_{F^{(s)}}}^{2},
\label{ObsGammaredwidth1}
\end{align}
where the observable and formal reduced widths ${\tilde \gamma}_{\tau\,j_{i}l_{i}J_{F^{(s)}}}^{2}$ and $\gamma_{\tau\,j_{i}l_{i}J_{F^{(s)}}}^{2}$, correspondingly, are related by
\begin{align}
{\tilde \gamma}_{\tau\,j_{i}l_{i}J_{F^{(s)}}}^{2}= \frac{\gamma_{\tau\,j_{i}l_{i}J_{F^{(s)}}}^{2}   }{ 1+ \gamma_{\tau\,j_{i}l_{i}J_{F^{(s)}}}^{2} [{\rm d}S_{l_{i}}(E_{xA})/{\rm d}E_{xA}] \big|_{E_{xA}= E_{\tau} }}. 
\label{obsformwidth1}
\end{align}
$E_{1} = -\varepsilon_{xA}^{(s)}$ and $E_{2} = E_{R}$, $E_{R}$ is the resonance energy corresponding to the level $\tau=2$.

For the subthreshold resonance ($\tau=1$) \cite{muk99}
\begin{widetext}
\begin{align}
\frac{[C_{1\,j_{i}l_{i}J_{F^{(s)}}}]^{2}\,W_{-\,\eta_{xA}^{(s)},l_{i}+1/2}^{2}(  2\,\kappa_{xA}^{(s)}\,R_{xA}) }{2\,\mu_{xA}\,R_{xA}}=  \frac{   \gamma_{1\,j_{i}l_{i}J_{F^{(s)}}}^{2}   }{ 1+ \gamma_{1\,j_{i}l_{i}J_{F^{(s)}}}^{2} [{\rm d}S_{l_{i}}(E_{xA})/{\rm d}E_{xA}] \big|_{E_{xA}= -\varepsilon_{xA}^{(s)} }} = {\tilde \gamma}_{1\,j_{i}l_{i}J_{F^{(s)}}}^{2},
\label{ANCredwidth1}
\end{align}
 \end{widetext}
where $ {\tilde \gamma}_{1\,j_{i}l_{i}J_{F^{(s)}}}^{2}$ and $\gamma_{1\,j_{i}l_{i}J_{F^{(s)}}}^{2}$ are the observed and formal reduced widths of the subthreshold resonance, $\,C_{1\,j_{i}l_{i}J_{F^{(s)}}}\,$ is the asymptotic normalization coefficient (ANC) of the subthreshold bound state $(x\,A)^{(s)}$ for the decay to the channel $(x+A)_{1\,j_{i}l_{i}J_{F^{(s)}}}$, $\,W_{-\eta_{xA}^{(s)},l_{i}+1/2}(\,2\,\kappa_{xA}^{(s)}\,R_{xA})$ is the Whittaker function, $\eta_{xA}^{(s)}= (Z_{x}\,Z_{A}/137)\mu_{xA}/\kappa_{xA}^{(s)}$  and $\kappa_{xA}^{(s)}$ are the $x-A$ Coulomb parameter and the bound-state wave number of the subthreshold bound state $\,F^{(s)}$, $\,\mu_{xA}$ is the reduced mass of $x$ and $A$, $\,Z_{j}\,e$ is the charge of nucleus $j$, $\,S_{l_{i}}(E_{xA})$ is the $R$-matrix Thomas shift function.

Now we return to the transfer reaction amplitude $M_{M_{a}\,M_{A}\,\tau}^{M_{s}\,M_{F_{\tau}^{(s)}}}({\rm {\bf k}}_{sF_{\tau}},{\rm {\bf k}}_{aA})$.
To calculate it we use the three-body approach   in which we neglect the internal degrees of freedom of particles $x,\,A$ and $s$.  The potential ${\overline V}_{xA}(r_{xA})$ depends only on the distance between $x$ and $A$.  Then the amplitude of the direct  transfer reaction (\ref{transfreact1}) in the plane-wave, surface-integral approximation reduces to \cite{reviewpaper,muk2011}
\begin{widetext}
\begin{align}
&M_{M_{a}\,M_{A}\,\tau}^{M_{s}\,M_{F_{\tau}^{(s)}}}({\rm {\bf k}}_{sF_{\tau}},{\rm {\bf k}}_{aA}) =  \frac{\sqrt{\pi}}{\mu_{xA}}\,i^{l_{i}}\,\varphi_{sx}(p_{sx})\,R_{xA}\,\Upsilon_{\tau\,j_{i}l_{i}J_{F^{(s)}}}(R_{xA})\,
{\tilde M}_{l_{i}}\sum\limits_{M_{x}\,m_{j_{i}}\,m_{l_{i}}}\, <J_{s}\,M_{s}\,\,J_{x}\,M_{x} \big|J_{a}\,M_{a}>         \nonumber\\
& \times <J_{x}\,M_{x}\,\,J_{A}\,M_{A} \big|j_{i}\,m_{j_{i}}> \,<j_{i}\,m_{j_{i}}\,\,l_{i}\,m_{l_{i}} \big| J_{F^{(s)}}\,M_{  F_{\tau}^{ (s) } }>Y_{l_{i}\,m_{l_{i}}}^{*}({\rm {\bf {\hat p}}}_{xA})
\nonumber\\
&=\frac{\sqrt{\pi}}{\mu_{xA}}\,i^{l_{i}}\,\varphi_{sx}(p_{sx})\,\sqrt{2\,\mu_{xA}\,R_{xA}}\,\gamma_{\tau\,j_{i}l_{i}J_{F^{(s)}}}
{\tilde M}_{l_{i}}\,\sum\limits_{M_{x}\,m_{j_{i}}\,m_{l_{i}}}\, <J_{s}\,M_{s}\,\,J_{x}\,M_{x} \big|J_{a}\,M_{a}>         \nonumber\\
& \times <J_{x}\,M_{x}\,\,J_{A}\,M_{A} \big|j_{i}\,m_{j_{i}}> \,<j_{i}\,m_{j_{i}}\,\,l_{i}\,m_{l_{i}} \big| J_{F^{(s)}}\,M_{F_{\tau}^{(s)}}>\,Y_{l_{i}\,m_{l_{i}}}^{*}({\rm {\bf {\hat p}}}_{xA}),
\label{THradamplSurf11}
\end{align}
\end{widetext}
\begin{widetext}
\begin{align}
&{\tilde M}_{l_{i}} =\Bigg\{ j_{l_{i}}(p_{xA}\,R_{xA}) \Big[ B_{l_{i}}(k_{xA},\,R_{xA}) - 1 - D_{l_{i}}(p_{xA},\,R_{xA}) \Big]
+ 2\mu_{xA}\frac{Z_{x}Z_{A}}{137}\,\int\limits_{R_{xA}}^{\infty}{\rm d}r_{xA}\,  j_{l_{i}}(p_{xA}\,r_{xA})   \frac{O_{l_{i}}(r_{xA})}{O_{l_{i}}(R_{xA})} \Bigg\},
\label{Mtau3}
\end{align}
\begin{eqnarray}
D_{l_{i}}(p_{xA},\,R_{xA})= R_{xA}\,\frac{  {\partial \ln j_{l_{i}}(p_{xA},r_{xA} ) }}{\partial r_{xA} } \Big|_{r_{xA}=R_{xA}},\,\,
B_{l_{i}}(k_{xA},\,R_{xA})= R_{xA}\,\frac{ {\partial \ln O_{l_{i}}(k_{xA},r_{xA})}}{\partial r_{xA}}\Big|_{r_{xA}=R_{xA}}.
\label{DB1}
\end{eqnarray}
\end{widetext}
Here, $\varphi_{sx}(p_{sx})$ is the Fourier transform of the radial part of the $\,s$-wave bound-state wave function $\varphi_{sx}(p_{sx})$ of the $a=(s\,x)$.  Also, $\kappa_{sx}=\sqrt{ 2\,\mu_{sx}\,\varepsilon_{sx}}\,$ is the wave number of the bound-state $\,a=(s\,x)$, $\,\varepsilon_{sx}\,$ is its binding energy for the virtual decay  $a \to s+ x$. Since  particles $s$ and $x$  are structureless,  the spectroscopic factor of the bound state $a=(s\,x)$ is unity and we can use just the bound-state wave function $\varphi_{sx}$. In the center-off-mass of the reaction (\ref{reaction1})  ${\rm {\bf k}}_{aA}= {\rm {\bf k}}_{a}$, $\,\,{\rm {\bf k}}_{sF_{\tau}}={\rm {\bf k}}_{s}$ and  
\begin{align}
{\rm {\bf p}}_{xA}=  {\rm {\bf  k}}_{a} - \frac{m_{A}}{m_{F}}\,{\rm {\bf k}}_{s},    \qquad
{\rm {\bf p}}_{sx}= {\rm {\bf k}}_{s} - \frac{m_{s}}{m_{a}}\,{\rm {\bf k}}_{a}
\label{pxA1psx1}
\end{align}
are the off-shell $x-A$ and $s-x$ relative momenta in the vertices $x+ A \to F_{\tau}$ and $a \to s+x$ of the  diagram in Fig. \ref{fig:fig_polediagram1}, correspondingly,
 $\,{\rm {\bf p}}_{x} = {\rm {\bf k}}_{a} - {\rm {\bf k}}_{s}$  is the off-shell momentum of the transferred virtual particle $\,x$, $\,{\rm {\bf k}}_{j}$ is the on-shell momentum of particle $j$.  Also $\,k_{s}$ and $E_{xA}$ are related by the energy conservation:
\begin{align}
E_{aA} - \varepsilon_{sx} = E_{xA} + k_{s}^{2}/(2\,\mu_{sF}), 
\label{energconserv1}
\end{align}
where $\mu_{sF}$ is the reduced mass of particles $s$ and $F$.
The amplitude $\,M_{M_{a}\,M_{A}\,\tau}^{M_{s}\,M_{F_{\tau}^{(s)}}}$ is taken at fixed projections of the spins of the entry  and the exit  particles of the reaction $a+ A \to s + F_{\tau}$,   $\,\,\tau=1,\,2$.

Now we consider the amplitude  $V_{\nu},\,\,\nu=1,2,$ describing the radiative decay of the intermediate resonance $F_{\nu} \to F + \gamma$:
\begin{widetext}
\begin{align}
&V_{M_{F_{\nu}}^{(s)}\,\nu}^{M_{F}\,M\,\lambda}= -\int\,{\rm d}{\rm {\bf r}}_{xA}
<I_{xA}^{F}({\rm {\bf r}}_{xA})\,\big|{\rm {\bf {\hat J}}}({\rm {\bf r}}) \big|\Upsilon_{\nu}({\rm {\bf r}}_{xA})>\cdot {\rm {\bf A}}^{*}_{\lambda\,{\rm {\bf k}}_{\gamma}}({\rm {\bf r}}),
\label{Vnu1}
\end{align}
\end{widetext}
where  $I_{xA}^{F}({\rm {\bf r}}_{xA})$ is the overlap function of the bound-state wave functions of $\,x,$ $\,A$ and the ground state of $\,F=(x\,A)$. Again, for the point-like nuclei $x$ and $A$  the overlap function $I_{xA}^{F}({\rm {\bf r}}_{xA})$  can be replaced by the single-particle  bound-state wave function of $(xA)$ in the ground state. Also ${\rm {\bf A}}^{*}_{\lambda\,{\rm {\bf k}}_{\gamma}}({\rm {\bf r}}) $ is the electromagnetic  vector potential of the photon with helicity $\lambda=\pm 1$ and momentum ${\rm {\bf k}}_{\gamma}$ at coordinate ${\rm {\bf r}}_{xA}$.
$\,{\rm {\bf {\hat J}}}({\rm {\bf r}})$  is the charge current density operator. Matrix element in Eq. (\ref{Vnu1}) is written assuming that on the first stage of the reaction the excited state $F_{\nu},\,\nu=1,2,$ is populated, which subsequently decays to the ground state $F$.

Using the multipole expansion of  the vector potential and leaving only
the electric components with the lowest allowed multipolarities $L$ and using the long wavelength approximation for ${\rm {\bf {\hat J}}}({\rm {\bf r}}) $, see for details \cite{muk2016},  we get
\begin{widetext}
\begin{align}
& V_{M_{F_{\nu}^{(s)}}\,\nu}^{M_{F}\,M\,\lambda}  = -\frac{1}{2\,\pi}\,\sum\limits_{L}\,\sqrt{ \frac{1}{2\,k_{\gamma}}} \,\sqrt{\frac{L+1}{L}}\,\sqrt{{\hat J}_{F}\,{\hat l}_{f}}\,\frac{i^{-L}\,k_{\gamma}^{L}}{(2\,L-1)!!}\,e\,Z_{eff{(L)}}\,\big[D^{L}_{M\,\lambda}(\phi,\,\theta,0)\big]^{*}\,   \nonumber\\
& \times\,<l_{f}\,0\,\,L\,0\big|l_{i}\,0>\,(-1)^{l_{i} - j_{i} -J_{F^{(s)}}^{L} }\,<J_{F}\,M_{F}\,\,L\,M \big|J_{F^{(s)}}^{L}\,M_{F_{\nu}^{(s)}}^{L}>\,\Bigg\lbrace \begin{array}{ccc} l_{f} \,\,j_{i}\,\,J_{F} \\
 J_{F^{(s)}}^{L}\,L\,l_{i}  \end{array} \Bigg\rbrace\,R_{\nu\,j_{f}\,l_{f}\,J_{F}\,j_{i}\,l_{i}\,J_{F^{(s)}}^{L}}^{L}   \nonumber\\
&= \,\frac{\sqrt{2}}{4\,\pi}\,\sum\limits_{L}i^{-L}\,(-1)^{L+1}\sqrt{\hat L} \,k_{\gamma}^{L-1/2} \,[\gamma_{(\gamma)\,\nu\,J_{F}\,L}^{J_{F^{(s)}}^{L}} ]\, \big[D^{L}_{M\,\lambda}(\phi,\,\theta,0)\big]^{*}\,<J_{F}\,M_{F}\,\,L\,M \big|J_{F^{(s)}}^{L}\,M_{F_{\nu}^{(s)}}^{L}>,
 \label{VMFgamma2}
\end{align}
\end{widetext}
where $\,\gamma_{(\gamma)\,\nu\,J_{F}\,L}^{J_{F^{(s)}}^{L}}$ is the formal $R$-matrix radiative width amplitude for the electric  $\,EL$ transition $\,J_{F^{(s)}}^{L} \to J_{F}$  given by the sum of the internal and external radiative width amplitudes, see Eqs (32) and (33) from \cite{muk2017},  in  which we singled out $\sqrt{2}\,k_{\gamma}^{L+1/2}$. Because now we take into account a few multipolarities $L$, we replace the previously introduced spin of the intermediate resonance $J_{F^{(s)}}$ by $J_{F^{(s)}}^{L}$, where the superscript $L$ denotes the multipolarity of the $EL$ transition to the ground state $F$. Replacement of $J_{F^{(s)}}$ by $J_{F^{(s)}}^{L}$ takes into account that the spins of the intermediate excited states are different for different multipolarities. Since we added the subscript $L$ to the spin of the intermediate resonance we added the same subscript to its projection $M_{F_{\nu}^{(s)}}^{L}$.

 The determined  radiative width amplitude is related to the formal resonance radiative width by the standard equation
\begin{align}
\Gamma_{(\gamma)\,\nu\,J_{F}\,L}^{J_{F^{(s)}}^{L}}= 2\,k_{\gamma}^{L+1/2}\,(\gamma_{(\gamma)\,\nu\,J_{F}\,L}^{J_{F^{(s)}}^{L}})^{2}.
\label{Gg1}
\end{align}
Note that the observable radiative width is related to the formal one by
\begin{align}
({\tilde \gamma}_{(\gamma)\,\nu\,J_{F}\,L}^{J_{F^{(s)}}^{L}})^{2} = \frac{(\gamma_{(\gamma)\,\nu\,J_{F}\,L}^{J_{F^{(s)}}^{L}})^{2}}{ 1+ \gamma_{\nu\,j_{i}l_{i}J_{F^{(s)}}}^{2} [{\rm d}S_{l_{i}}(E_{xA})/{\rm d}E_{xA}] \big|_{E_{xA}= E_{\nu}}}.
\label{observformradwidth1}
\end{align}
We consider two-level approach with $\nu=1$ corresponding to the subthreshold resonance and $\nu=2$
to the resonance at $E_{xA}>0$. Then $E_{\nu}=-\varepsilon_{xA}^{(s)}$ for $\nu=1$  and  $E_{\nu}= E_{R}$ for $\nu=2$ with $E_{R}$ being the resonance energy corresponding to the level
$\nu=2$.
This observable radiative width is related to the observable resonance radiative width as
\begin{align}
{\tilde \Gamma}_{(\gamma)\,\nu\,J_{F}\,L}^{J_{F^{(s)}}^{L}}= 2\,k_{\gamma}^{L+1/2}\,({\tilde \gamma}_{(\gamma)\,\nu\,J_{F}\,L}^{J_{F^{(s)}}^{L}})^{2}.
\label{Gg2}
\end{align}

Also in Eq. (\ref{VMFgamma2})  $\,M$  is  the projection  of the angular momentum $L$ of the emitted photon (multipolarity of the electromagnetic transition),  $\,e\,Z_{eff(L)}$ is the effective charge of the $x+A$ system for the electric transition $EL$. The matrix element  $R_{\nu j_{f}\,l_{f}\,J_{F}\,j_{i}\,l_{i}\,J_{F^{(s)}}^{L}}^{L} $ is
\begin{align}
&R_{\nu j_{f}\,l_{f}\,J_{F}\,j_{i}\,l_{i}\,J_{F^{(s)}}^{L}}^{L} = <r_{xA}^{L+2}  I_{j_{i}\,l_{f}\,J_{F}}(r_{xA}) \Upsilon_{\nu\,j_{i}\,l_{i}\,J_{F^{(s)}}^{L}}(r_{xA})>.
\label{matrelem1}
\end{align}
$\Upsilon_{\nu\,j_{i}\,l_{i}\,J_{F^{(s)}}^{L}}(r_{xA})$ is the resonant scattering wave function in the $R$-matrix approach  whose  external part is given by Eq. (\ref{extreswf1}) and the internal
resonant wave function $X_{int\,\tau}$ in the $R$-matrix approach matches the external one on the border $r_{xA}=R_{xA}$ and satisfies the boundary condition
\begin{align}
X_{int\,\tau}(k_{xA},\,R_{xA})=  \sqrt{2\,\mu_{xA}\,R_{xA}}\,\gamma_{\tau\,j_{i}l_{i}J_{F^{(s)}}^{L}}.
\label{Xintext1}
\end{align}
 For $\,\tau=1$ $\,X_{int\,1}$ is the overlap function of the bound-state wave functions of $F^{(s)}=(x\,A)^{(s)}$, $\,x$  and $\,A$, which is normalized to untiy over the internal region $r_{xA} \leq R_{xA}$.

Substituting Eqs. (\ref{THradamplSurf11}) and (\ref{VMFgamma2})  into Eq.  (\ref{THradamplfull1}) we get  the expression for the indirect  reaction amplitude
\begin{widetext}
\begin{align}
&M_{M_{a}\,M_{A}}^{M_{s}\,\,M_{F}\,M\,\lambda} = \,\frac{\varphi_{sx}(p_{sx})}{2}\,\sqrt{\frac{R_{xA}}{\pi\,\mu_{xA}}}\,\sum\limits_{L}\,(-1)^{L+1}\,{\hat L}^{1/2}\,k_{\gamma}^{L-1/2}\,\big[D^{L}_{M\,\lambda}(\phi,\,\theta,0)\big]^{*}\,\sum\limits_{l_{i}}\,i^{l_{i} -L}\,{\tilde M}_{l_{i}} \nonumber\\
& \times\,  \sum\limits_{\nu,\,\tau=1}^{2}\,\gamma_{(\gamma)\,\nu\,J_{F}\,L}^{J_{F^{(s)}}^{L}}\,{\rm {\bf A}}_{\nu\,\tau}^{L} \,\gamma_{\tau\,j_{i}l_{i}J_{F^{(s)}}^{L}}\,\sum\limits_{ M_{F_{\nu}^{(s)}}^{L} }\, <J_{F}\,M_{F}\,\,L\,M \big|J_{F^{(s)}}^{L}\,M_{F_{\nu}^{(s)}}^{L}>                                                           \nonumber\\
& \times  \sum\limits_{m_{j_{i}}\,m_{l_{i}}\,M_{x} }<j_{i}\,m_{j_{i}}\,\,l_{i}\,m_{l_{i}}\big| J_{F^{(s)}}^{L}\,M_{F_{\tau}^{(s)}}^{L}>\, <J_{x}\,M_{x}\,\,J_{s}\,M_{s} \big|J_{a}\,M_{a}>\,<J_{x}\,M_{x}\,\,J_{A}\,M_{A} \big| j_{i}\,m_{j_{i}}>\,Y_{l_{i}\,m_{l_{i}}}^{*} ({\rm {\bf {\hat p}}}_{xA}).
\label{MTHtot1}
\end{align}
\end{widetext}

The amplitude $M_{M_{a}\,M_{A}}^{M_{s}\,\,M_{F}\,M\,\lambda}$  describes the indirect reaction proceeding through the intermediate resonances, which decay to the ground state $F=(x\,A)$ by emitting  photons. Equation (\ref{MTHtot1}) is generalization of Eq. (\ref{THradamplfull1})
by including the sum over multipolarities $L$ corresponding to the radiative electric transitions from the intermediate resonances with the spins $J_{F^{(s)}}^{L}$ to the ground state $F$ with the spin $J_{F}$. Note also that we assume that each transition of multipole $L$ is contributed by two levels. It requires the two-level generalized $R$-matrix approach. 
The generalization of Eq. (\ref{MTHtot1}) for three- or more-level cases is straightforward. 
In Eq. (\ref{MTHtot1}) the reaction part and radiative parts are interconnected by the $R$-matrix level matrix elements ${\rm {\bf A}}_{\nu\,\tau}^{L}$.

The part $\sum\limits_{\nu,\,\tau=1}^{2}\,\gamma_{(\gamma)\,\nu\,J_{F}\,L}^{J_{F^{(s)}}^{L}}\,{\rm {\bf A}}_{\nu\,\tau}^{L} \,\gamma_{\tau\,j_{i}l_{i}J_{F^{(s)}}^{L}}$ is the standard $R$-matrix term for the binary resonant radiative capture reaction. However, we analyze the three-body reaction  $a(x\,s) + A \to s +F +\gamma$ with the spectator $s$ in the final state rather than the standard two-body radiative capture reaction $x +A \to F+ \gamma$. This difference leads to the 
generalization of the standard $R$-matrix approach for the three-body reactions resulting
in the appearance of the additional terms, $\varphi_{sx}(p_{sx})\,{\tilde M}_{l_{i}}$. That is why we call the developed approach the generalized $R$-matrix method for the indirect resonant radiative capture reactions. 

(i) The most important feature of this approach is that the indirect reaction amplitude does not contain 
the penetrability factor $P_{l_{i}}(E_{xA},R_{xA})$ in the entry channel of the sub-reaction (\ref{reaction1}). This factor is the main obstacle to measure the astrophysical factor of this reaction if one uses direct measurements. 
The absence of this penetrability factor in the entry channel of the sub-reaction allows one to use the indirect method to get the information about the astrophysical factor of the sub-reaction.\\
(ii) The indirect reaction amplitude is parameterized in terms of the formal $R$-matrix  width amplitudes, which are connected to the observable resonance widths.\\
(iii) A problem with the relative phase of the interfering multipoles does not exist anymore
because the R-matrix phase factor appearing in $M_{M_{a}\,M_{A}\,\tau}^{M_{s}\,M_{F_{\tau}^{(s)}}}$ is compensated by the complex conjugated phase factor in $V_{M_{F_{\nu}^{(s)}}\,\nu}^{M_{F}\,M\,\lambda}$. 

We take the indirect reaction amplitude at fixed projections of the spins of the initial and final particles including the fixed projection $M$ of the orbital momentum $L$ of the emitted photon and fixed its chirality $\lambda$.  For example, for the ${}^{12}{\rm C}(\alpha,\,\gamma){}^{16}{\rm O}$ reaction the electric dipole $\,E1$ ($L=1$) and quadrupole $\,E2$ ($L=2$) transitions do contribute and they interfere. In the long wavelength approximation only minimal allowed $l_{i}$ for given $L$ does contribute. For example, for the case considered below $\,l_{f}=0$ $\,\,l_{i}=L=1$ for the dipole  and $\,l_{i}=L=2$ for the quadrupole electric transitions. The dimension of the $R$-matrix level matrix ${\rm {\bf A}}^{L} $ depends on the number of the levels taken into account for each $L$. 

The indirect reaction amplitude  depends on the off-shell momenta ${\rm {\bf p}}_{sx}$ and 
${\rm {\bf p}}_{xA}$. Both off-shell momenta are expressed in terms of ${\rm {\bf k}}_{a}$ and ${\rm {\bf k}}_{s}$, see Eq. (\ref{pxA1psx1}). Also the the indirect reaction amplitude depends on the momentum of the emitted photon ${\rm {\bf k}}_{\gamma}$  whose direction is determined by the angles in the Wigner $D$-function. 
 In the center-off-mass of the  reaction 
(\ref{THMreaction1}) from the energy conservation we get
\begin{align}
E_{aA} + Q = E_{sF} + k_{\gamma}, \qquad  k_{\gamma} = E_{xA} + \varepsilon_{xA}, 
\label{energconserv1}
\end{align}
where $\,E_{sF}= k_{s}^{2}/(2\,\mu_{sF})$ and $\,Q=\varepsilon_{xA} - \varepsilon_{sx}$. We neglect the recoil of nucleus $F$ during the photon emission. Thus $k_{\gamma}$ can be expressed in terms of  $k_{s}$ while $k_{\gamma}$ and $E_{xA}$ are related.
The expression for $p_{xA}$ is needed to calculate ${\tilde M}_{l_{i}}$. From the energy-momentum conservation law in the three-ray vertices $a \to s+x$ and $ x+ A \to F^{(s)}$ of the diagram in Fig. \ref{fig:fig_polediagram1} we get \cite{reviewpaper}
\begin{align}
E_{xA}= \frac{p_{xA}^{2}}{2\,\mu_{xA}} - \frac{p_{sx}^{2}}{2\,\mu_{sx}} -\varepsilon_{s\,x}.
\label{ExApxA1}
\end{align}

\section{ Differential cross sections}
\label{difcrsct1}

\subsection{Triple differential cross section}

Let us consider the indirect resonant reaction contributed by different interfering multipoles $L$. For each $L$ we assume two-level contribution.
Then the triple differential cross section of the resonant indirect radiative capture reaction for unpolarized initial and final particles (including the photon) in the center-off-mass of the reaction (\ref{THMreaction1}) is  given by
\begin{widetext}
\begin{align}
&\frac{ {\rm d}\sigma}{  {\rm d}{\Omega}_{  {\rm{\bf {\hat k}}}_{s} }\,{\rm d}{\Omega}_{{\rm {\bf {\hat k}}}_{\gamma}}\, {\rm d}{E_{sF}} }=
 \frac{\mu_{aA}\,\mu_{sF}}{{\hat J}_{a}\,{\hat J}_{A}(2\,\pi)^{5}}\,\frac{k_{sF}\,k_{\gamma}^{2}}{k_{aA}}
\sum\limits_{M_{a}\,M_{A}\,M_{s}\,M_{F}\,M\,\lambda}\Big|M_{M_{a}\,M_{A}}^{M_{s}\,M_{F}\,M\,\lambda}\Big|^{2}
 \nonumber\\
&= - \frac{1}{(2\,\pi)^{7} }\,\frac{\mu_{aA}\,\mu_{sF}}{{\hat J}_{x}\,{\hat J}_{A}} \,\frac{\varphi_{sx}^{2}(p_{sx})\,R_{xA}}{4\,\mu_{xA}}\,\frac{ k_{sF}}{k_{aA}}\,(-1)^{J_{F}  - j_{i} }\,
  \sum\limits_{L'\,L} (-1)^{L'+L}\, k_{\gamma}^{L'+L+1}\,\,{\hat J}_{F^{(s)}}^{L'}\,{\hat J}_{F^{(s)}}^{L}\,\sqrt{ {\hat L}'\,{\hat L}}\,  
\sum\limits_{l_{i}'\,l_{i}\,l}i^{L'-l_{i}'- L+l_{i}}\,\sqrt{ {\hat l}_{i}'\,{\hat l}_{i} }      \nonumber\\
& \times\,  {\tilde M}_{l_{i}'}^{*}\,{\tilde M}_{l_{i}}\, \Bigg\lbrace \begin{array}{ccc} j_{i}\,l_{i}'\, J_{F^{(s)}}^{L'} \\  l\,J_{F^{(s)}}^{L}\,l_{i} \end{array} \Bigg\rbrace\,\Bigg\lbrace \begin{array}{ccc} J_{F^{(s)}}^{L'}\, J_{F}\,L' \\
L\,l\,J_{F^{(s)}}^{L} \end{array} \Bigg\rbrace   \sum\limits_{\nu', \,\nu,\,\tau',\, \tau=1}^{2}\,\big[\gamma_{ (\gamma)\,\nu'\,J_{F}\,L'}^{J_{F^{(s)}}^{L'}}  \big]^{*} [\gamma_{(\gamma)\,\nu\,J_{F}\,L}^{J_{F^{(s)}}^{L}} ]\,\big[{\bf A}_{v'\,\tau'}^{L'} \big]^{*}\,\big[{\bf A}_{\nu\,\tau}^{L} \big]\     \nonumber\\
& \times ,\gamma_{\tau'\,j_{i}l_{i}'J_{F^{(s)}}^{L'} }\,\gamma_{\tau\,j_{i}l_{i}J_{F^{(s)}}^{L}} \, <l_{i}'\,0\,\,l_{i}\,0\big|l\,0>\,<L'\,1\,\,L\,-1 \big| l\,0>\,[1+ (-1)^{L'+L+l}]\,P_{l}(cos{\theta}).
\label{tripplediffcrsect1}
\end{align}
\end{widetext}
To obtain Eq. (\ref{tripplediffcrsect1}) we adopted $z\, ||\, {\rm {\bf {\hat  p}}}_{xA} $,  that is,  
$Y_{l\,m_{l}}({\rm {\bf {\hat  p}}}_{xA}) = \sqrt{\frac{{\hat l}}{4\,\pi}}\,\delta_{m_{l}\,0}$.
Thus, in the plane-wave approximation the direction ${\rm {\bf {\hat  p}}}_{xA}$ becomes the axis of the symmetry.
Note that if we replace the plane waves by the distorted waves the vestige of this symmetry will still survive.

We remind that the radiative transition  $J_{F^{(s)}}^{L} \to J_{F}$ is the electric $EL$ where $J_{F^{(s)}}^{L}$ is the spin of the intermediate state (subthreshold resonance or resonance).

For a more simple case when only one multipole $L$ contributes into the radiative transition, the triple differential cross section 
takes the form:
\begin{widetext}
\begin{align}
&\frac{ {\rm d}\sigma}{  {\rm d}{\Omega}_{  {\rm{\bf {\hat k}}}_{s} }\,{\rm d}{\Omega}_{{\rm {\bf {\hat k}}}_{\gamma}}\, {\rm d}{E_{sF}} }= - \frac{1}{(2\,\pi)^{7} }\,\frac{\mu_{aA}\,\mu_{sF}}{{\hat J}_{x}\,{\hat J}_{A}} \,\frac{\varphi_{sx}^{2}(p_{sx})\,R_{xA}}{2\,\mu_{xA}}\,\frac{ k_{sF}}{k_{aA}}\,
 k_{\gamma}^{2\,L+1}\,(-1)^{J_{F}  - j_{i} }\,{\hat L}\,({\hat J}_{F^{(s)}}^{L})^{2}\,
\sum\limits_{l_{i}\,l}\,{\hat l}_{i}\,                                   \nonumber\\
& \times \big|{\tilde M}_{l_{i}}\,\big|^{2}\,\Bigg\lbrace \begin{array}{ccc} j_{i}\,l_{i}\, J_{F^{(s)}}^{L} \\  l\,J_{F^{(s)}}^{L}\,l_{i} \end{array} \Bigg\rbrace\,\Bigg\lbrace \begin{array}{ccc} J_{F^{(s)}}^{L}\, J_{F}\,L \\
L\,l\,J_{F^{(s)}}^{L} \end{array} \Bigg\rbrace   \sum\limits_{\nu', \,\nu,\,\tau',\, \tau=1}^{2}\,\big[\gamma_{ (\gamma)\,\nu'\,J_{F}\,L}^{J_{F^{(s)}}^{L}}  \big]^{*} [\gamma_{(\gamma)\,\nu\,J_{F}\,L}^{J_{F^{(s)}}^{L}} ]\,\big[{\bf A}_{v'\,\tau'}^{L} \big]^{*}\,\big[{\bf A}_{\nu\,\tau}^{L} \big]\     \nonumber\\
& \times ,\gamma_{\tau'\,j_{i}l_{i}J_{F^{(s)}}^{L}}\,\gamma_{\tau\,j_{i}l_{i}J_{F^{(s)}}^{L}} \, <l_{i}\,0\,\,l_{i}\,0\big|l\,0>\,<L\,1\,\,L\,-1 \big| l\,0>\,P_{l}(cos{\theta}).
\label{tripplediffcrsect2}
\end{align}
\end{widetext}
Also formally we keep the summation over $l_{i}$, in the long wavelength approximation for given $L$ at astrophysically relevant energies only minimal allowed $l_{i}$ does contribute. 

The triple differential cross section depends on ${\rm {\bf k}}_{s}$ and $\,{\rm {\bf k}}_{\gamma}$. Because we neglected the recoil of the final nucleus $F$, $\,k_{s}$ and $k_{\gamma}$ are related by Eq. (\ref{energconserv1}). We remind that we selected axis $z||{\rm{\bf p}}_{xA}$.  Hence the photon's scattering angle is counted from ${\rm{\bf p}}_{xA}$, which itself is determined by ${\rm {\bf k}}_{s}$.
 Thus the angular dependence of the triple differential cross section
actually determines the angular correlation between the emitted photons from the intermediate excited state $F^{*}$ and the spectator $s$. Because we consider the three-body reaction (\ref{THMreaction1})
the angular correlation function depends also on the spins $J_{F^{(s)}}^{L}$ of the intermediate nucleus $F^{*}$ which decays to $F$.

By choosing QF kinematics, $p_{sx}=0$, one can provide the maximum of the triple differential cross section due to the maximum of $\varphi_{sx}^{2}(p_{sx})$. At fixed ${\rm {\bf k}}_{s}$ the triple differential cross section determines the emitted photon's angular distribution, which is contributed by different interfering multipoles $L$.
By measuring the photon's angular distributions at different photon's energies (that is, at different $k_{s}$ or $E_{xA}$) one can determine the energy dependence of the photon's angular distribution. However, a wide variation of ${\rm {\bf k}}_{s}$ 
away from the QF kinematics ${\rm {\bf p}}_{sx} = {\rm {\bf k}}_{s} - (m_{s}/m_{a})\,{\rm {\bf k}}_{a}=0$ will decrease the differential cross section due to the drop of $\varphi_{sx}^{2}(p_{sx})$. Usually, in indirect methods $\,{\rm {\bf k}}_{s}$ is varied in the interval in which $p_{sx} \leq \kappa_{sx}$ \cite{reviewpaper}. \\

\subsection{Double differential cross section}
Integrating the triple differential cross section over the the photon's solid angle ${\Omega}_{{\rm {\bf {\hat k}}}_{\gamma}}$ we get the non-coherent sum of the double differential cross sections with different multipoles $L$:
\begin{widetext}
\begin{align}
&\frac{ {\rm d}\sigma}{  {\rm d}{\Omega}_{  {\rm{\bf {\hat k}}}_{s} }\, {\rm d}{E_{sF}} }=
 \frac{1}{(2\,\pi)^{6} }\,\frac{\mu_{aA}\,\mu_{sF}}{{\hat J}_{x}\,{\hat J}_{A}} \,\frac{\varphi_{sx}^{2}(p_{sx})\,R_{xA}}{\mu_{xA}}\,\frac{ k_{sF}}{k_{aA}}\,
 \sum\limits_{L}\,\sqrt{{\hat L}\,{\hat J}_{F^{(s)}}^{L}}\,k_{\gamma}^{2L+1}\,\sum\limits_{l_{i}}\,\big| {\tilde M}_{l_{i}}\big|^{2}                                  \nonumber\\
& \times\,\sum\limits_{\nu', \,\nu,\,\tau',\, \tau=1}^{2}\,\big[\gamma_{ (\gamma)\,\nu'\,J_{F}\,L}^{J_{F^{(s)}}^{L}}  \big]^{*} [\gamma_{(\gamma)\,\nu\,J_{F}\,L}^{J_{F^{(s)}}^{L}} ]\,\big[{\bf A}_{v'\,\tau'}^{L} \big]^{*}\,\big[{\bf A}_{\nu\,\tau}^{L} \big]\ \gamma_{\tau'\,j_{i}\,l_{i}\,J_{F^{(s)}}^{L}}\,\gamma_{\tau\,j_{i}\,l_{i}\,J_{F^{(s)}}^{L}}.
\label{doublediffcrsect11}
\end{align}
\end{widetext}
Despite of the virtual transferred particle $x$ in the diagram of Fig. \ref{fig:fig_polediagram1}, using the surface integral approach and generalized $R$-matrix we can rewrite the double differential cross section in terms of the 
on-the-energy-shell (OES) astrophysical factor for the resonant radiative capture $A(x,\,\gamma)F$ for the electric transition of the multipolarity $L$ and the relative orbital angular momentum $l_{i}$ of particles $x$ and $A$ in the entry channel of the $A(x,\,\gamma)F$ raiative capture. In the $R$-matrix formalism this astrophysical factor is given by
\begin{widetext}
\begin{align}
S_{EL,l_{i}}(E_{xA}) (MeVb) = 2\,\pi\,\lambda_{N}^{2}\,\frac{{\hat J}_{F^{(s)}}^{L}}{{\hat J}_{x}\,{\hat J}_{A}}\,\frac{1}{\mu_{xA}}\,m_{N}^{2}\,
e^{2\,\pi\,\eta_{i}}\,P_{l_{i}}(E_{xA},\,R_{xA})\,10^{-2}\,k_{\gamma}^{{\hat L}}\,\Big|\sum\limits_{\nu,\tau}[\gamma_{(\gamma)\,\nu\,J_{F}\,L}^{J_{F^{(s)}}^{L}} ]\,\big[{\bf A}_{\nu\,\tau}^{L} \big]\ \gamma_{\tau\,j_{i}\,l_{i}\,J_{F^{(s)}}^{L}}\Big|^{2}.
 \label{Sfactor1}
\end{align}   
\end{widetext}
Here, $\lambda_{N}=0.2118$ fm is the Compton nucleon wave length, $m_{N}= 931.5$ MeV is the atomic mass unit, $\mu_{xA}$ is the $x-A$ reduced mass expressed in MeV, $\eta_{i}$ is the $x-A$ Coulomb parameter at relative enerrgy $E_{xA}$.  
Then the indirect double differential cross section takes the form:
\begin{widetext}
\begin{align}
\frac{ {\rm d}\sigma}{  {\rm d}{\Omega}_{  {\rm{\bf {\hat k}}}_{sF} }\, {\rm d}{E_{sF}} }=KF\,\varphi_{sx}^{2}(p_{sx})\,R_{xA}
\,\sum\limits_{L} \,\sqrt{\frac{{\hat L}}{{\hat J}_{F^{(s)}}^{L}}}\,\sum\limits_{l_{i}}e^{-2\,\pi\,\eta_{i}}\,P_{l_{i}}^{-1}(E_{xA},\,R_{xA})\big| {\tilde M}_{l_{i}} \big|^{2}\,S_{EL,l_{i}}(E_{xA}),                                 
\label{doublediffcrsectSfctr1}
\end{align}
\end{widetext}
where 
\begin{align}
KF=  \frac{10^{2}}{(2\,\pi)^{7} }\,\frac{\mu_{aA}\,\mu_{sF}}{m_{N}^{2}\,\lambda_{N}^{2}}\,\frac{ k_{sF}}{k_{aA}}
\label{KF1}
\end{align}
is the kinematical factor.

Assume, for simplicity, that at higher energies only one $L$ and $l_{i}$ do contribute into Eq. (\ref{doublediffcrsectSfctr1}).
Then we can express the astrophysical factor measured in the indirect approach in terms of the indirect double cross section:
\begin{widetext}
\begin{align}
 S_{EL,l_{i}}(E_{xA})= \frac{ {\rm d}\sigma}{  {\rm d}{\Omega}_{  {\rm{\bf {\hat k}}}_{sF} }\, {\rm d}{E_{sF}} }\,\sqrt{\frac{ {\hat J}_{F^{(s)}}^{L} }{ {\hat L}}}  \,\frac{1}{KF\,\varphi_{sx}^{2}(p_{sx})\,R_{xA}}\,e^{2\,\pi\,\eta_{i}}\,P_{l_{i}}(E_{xA},\,R_{xA})\,
\big| {\tilde M}_{l_{i}} \big|^{-2}.
\label{Sfactrdblcrsect1}
\end{align}
\end{widetext}
As we have underscored, in the indirect method the absolute cross sections are not measured. 
Hence, the double differential cross section in Eq. (\ref{Sfactrdblcrsect1}) is not normalized.
However, if the astrophysical factor at higher energies is available from direct measurements, we can normalize the right-hand-side of Eq. (\ref{Sfactrdblcrsect1}):
\begin{widetext}
\begin{align}
 S_{EL,l_{i}}(E_{xA})= NF\,\frac{ {\rm d}\sigma}{  {\rm d}{\Omega}_{  {\rm{\bf {\hat k}}}_{sF} }\, {\rm d}{E_{sF}} }\,\sqrt{\frac{ {\hat J}_{F^{(s)}}^{L} }{ {\hat L}}}  \,\frac{1}{KF\,\varphi_{sx}^{2}(p_{sx})\,R_{xA}}\,e^{2\,\pi\,\eta_{i}}\,P_{l_{i}}(E_{xA},\,R_{xA})\,
\big| {\tilde M}_{l_{i}} \big|^{-2}.
\label{Sfactrdblcrsect11}
\end{align}
\end{widetext}
Here, $NF$ is an energy-independent normalization factor providing correct astrophysical factor $S_{EL,l_{i}}(E_{xA})$ at higher energies. 
Using this normalization factor we can determine with accuracy, which is not achievable in any direct approach, the astrophysical factors at energies $E_{xA} \to 0$. This is the main achievement of the indirect approach.
We remind that in our formalism we use the plane wave approximation rather than the distorted wave. But it should not affect the accuracy of our approach because the energy dependence of the transfer reaction cross section are similar for the distorted wave or the plane wave approach. The normalization factor
$NF$ compensates the inaccuracy of the plane wave approximation.

We summarize the technology of the indirect method to obtain the astrophysical factor.\\
(1) Measurements of the photon's angular distribution (photon-spectator angular correlation) at different $E_{xA}$ energies covering the interval from low energies relevant to nuclear astrophysics up to higher energy at which direct data are available. To cover a broad energy interval at fixed energy of the projectile the energy and scattering angle of the spectator should be varied near the QF kinematics ($p_{sx}=0$). \\
(2) Obtaining the indirect double differential cross section by integrating the triple differential cross section over the photon's scattering angle.\\
(3) Expressing the astrophysical factor in terms of the indirect double differential cross section.\\
(4) Normalization of astrophysical factor to the available experimental data at higher energy.\\
(5) Determination of the astrophysical factor at astrophysical energies. 
 
\section{Radiative capture ${}^{12}{\rm C}(\alpha,\,\gamma){}^{16}{\rm O}$ via  indirect reaction ${}^{12}{\rm C}({}^{6}{\rm Li},d\,\gamma){}^{16}{\rm O}$}

In this section we demonstrate the application of the developed formalism for the analysis
of the indirect reaction ${}^{12}{\rm C}({}^{6}{\rm Li},d\,\gamma){}^{16}{\rm O}$ to obtain the information about the astrophysical factor for the ${}^{12}{\rm C}(\alpha,\,\gamma){}^{16}{\rm O}$ at energies $< 1$ MeV. 

At low energies the astrophysical reaction under consideration is contributed by the $L=1$ and $L=2$ electric transitions \cite{barkerkajino,brune1999,kunz,assuncao,gai2015}. $E1$ transition to the ground state 
$J_{F}=0$ and $l_{f}=0$ proceeds as the resonant capture through the wing at $E_{\alpha\,{}^{12}{\rm C}}>0 $ of the subthreshold bound state $1^{-}$ at $E_{\alpha\,{}^{12}{\rm C}} = -0.045$ MeV, which works as the subthreshold resonance. Besides, the $E1$ transition to the ground state is contributed by the resonant capture through the low-energy tail of the $1^{-}$ resonance located at $E_{R}= 2.423$ MeV. The $E2$ transition is contributed by the subthreshold $2^{+}$ state at $E_{\alpha\,{}^{12}{\rm C}} = -0.2449$ MeV and low-energy tail of $2^{+}$ resonance at $2.68$ MeV.  

These four states are observable physical states contributing to the low-energy radiative capture under consideration. Besides these 
states, when fitting the data the artificial level was added for $E1$ transition (see, for example, 
\cite{barkerkajino,brune1999,kunz,assuncao,sayre} and references therein). In the present paper we calculate the  photon's angular distribution (the angular photon-deuteron correlation) at low energies down to the most effective astrophysical energy $E_{\alpha\,{}^{12}{\rm C}}=0.3$ MeV.
We take into account the mentioned four physical states and add one artificial state for the $E1$ transition.

The reduced widths of the subthreshold resonances are known from the experimental ANCs \cite{brune1999,avila} and the reduced width of the $1^{-},\,2.423$ MeV resonance is determined from the resonance width. 
We disregard the cascade transitions to the ground state of ${}^{16}{\rm O}$ through both subthreshold states because they are small. We also disregard the $E2$ direct radiative capture to the ground state which also interferes with the $E2$ radiative capture through the subthreshold resonance $2^{+}$. 

For the case under consideration $J_{x}=0,\,J_{A}=0,\,j_{i}=0,\,l_{i}=L=J_{F^{(s)}}^{L},\,J_{F}=0$ and the expression for the triple differential cross section for the case under consideration
simplifies to
\begin{widetext}
\begin{align}
&\frac{ {\rm d}\sigma}{  {\rm d}{\Omega}_{  {\rm{\bf {\hat k}}}_{s}}\,{\rm d}{\Omega}_{{\rm {\bf {\hat k}}}_{\gamma}}\, {\rm d}{E_{sF}} }
= - \frac{\mu_{aA}\,\mu_{sF}}{(2\,\pi)^{7} }\,\frac{\varphi_{sx}^{2}(p_{sx})\,R_{xA}}{2\,\mu_{xA}}\,\frac{ k_{sF}}{k_{aA}}\,\sum\limits_{L'\,L} (-1)^{L'+L}\, k_{\gamma}^{L'+L+1}\,
\sqrt{ {\hat L}'\,{\hat L}}                                   \nonumber\\
& \times\,  {\tilde M}_{L'}^{*}\,{\tilde M}_{L}\,\sum\limits_{\nu', \,\nu,\,\tau',\, \tau=1}^{2}\,\big[\gamma_{ (\gamma)\,\nu'\,0\,L'}^{L'}\,\big]^{*} [\gamma_{(\gamma)\,\nu\,0\,L}^{L} ]\,\big[{\bf A}_{v'\,\tau'}^{L'} \big]^{*}\,\big[{\bf A}_{\nu\,\tau}^{L} \big]\     \nonumber\\
& \times ,\gamma_{\tau'\,0\,L'\,L'} \,\gamma_{\tau\,0\,L\,L} \,\sum\limits_{l}\, <L'\,0\,\,L\,0\big|l\,0>\,<L'\,1\,\,L\,-1 \big| l\,0>\,P_{l}(cos{\theta}).
\label{tripplediffcrsect11}
\end{align}
\end{widetext}
Here, $a={}^{6}{\rm Li},\,A={}^{12}{\rm C},\,s=d,\,x=\alpha,\,F={}^{16}{\rm O}$.
This expression is used for the analysis of the indirect reaction ${}^{12}{\rm C}({}^{6}{\rm Li},d\,\gamma){}^{16}{\rm O}$ at low energies. We outline here some details of the calculations. 

After integration over the photon's solid angle we get the indirect double differential cross section:
\begin{widetext}
\begin{align}
\frac{ {\rm d}\sigma}{  {\rm d}{\Omega}_{  {\rm{\bf {\hat k}}}_{sF} }\,{\rm d}{E_{sF}} }=KF\,\varphi_{sx}^{2}(p_{sx})\,R_{xA}
\,\sum\limits_{L} \,\sqrt{\frac{{\hat L}}{{\hat J}_{F^{(s)}}^{L}}}\,e^{-2\,\pi\,\eta_{i}}\,P_{L}^{-1}(E_{xA},\,R_{xA})\big| {\tilde M}_{L}\big|^{2}\,S_{EL}(E_{xA}).                                
\label{doubledcrsct1}
\end{align}
\end{widetext}
Note that in the case under consideration $l_{i}=L$. 
Then at energies near the $1^{-}$ resonance at $2.423$ MeV where, as we will see below, the $E1$ transition completely dominates,
\begin{widetext}
\begin{align}
 S_{E1}(E_{xA})= NF\,\frac{ {\rm d}\sigma}{  {\rm d}{\Omega}_{  {\rm{\bf {\hat k}}}_{sF} }\, {\rm d}{E_{sF}} }\,\frac{1}{KF\,\varphi_{sx}^{2}(p_{sx})\,R_{xA}}\,e^{2\,\pi\,\eta_{i}}\,P_{1}(E_{xA},\,R_{xA})\,
\big| {\tilde M}_{1} \big|^{-2}.
\label{Sfactrdblcrsect412}
\end{align}
\end{widetext}
The $S(E1)$ astrophysical factor was measured at energies near $2.423$ MeV with a very good accuracy   \cite{redder,assuncao,Schuermann} and, should we have the experimental indirect double differential cross section measured expressed in arbitrary units, we can use Eq. (\ref{Sfactrdblcrsect412}) to normalize the $S_{E1}(E_{xA})$ to the experimental one at higher energies. After that, having 
measured indirect double differential cross section at $0.3$ MeV, we can determine the $S_{E1}(0.3 {\rm MeV}) + S_{E2}(0.3 {\rm MeV})$. 

In this paper we calculate  the photon's angular distribution at different $E_{\alpha\,{}^{12}{\rm C}}$ energies for the 
${}^{12}{\rm C}(\alpha,d\,\gamma){}^{16}{\rm O}$ reaction and how it is affected by the interference character (constructive or destructive) of the $1^{-}$ subthreshold bound state and $1^{-}$ resonance.  
The formal reduced width amplitude $\gamma_{1\,0\,1\,1}$ of the subthreshold bound state $1^{-}$ is related to the observable reduced width of this state
as
\begin{align}
(\gamma_{1\,0\,1\,1})^{2}=\frac{   ({\tilde \gamma}_{1\,0\,1\,1})^{2}  }{1 - ({\tilde \gamma}_{1\,0\,1\,1})^{2}\,\frac{ {\rm d}S_{1}(E_{xA}) }{  {\rm d}E_{xA} }\Big|_{ E_{xA}=-\varepsilon_{xA(1)}^{(s)}}},
\label{gamma111}
\end{align}
where $\varepsilon_{xA(1)}^{(s)} = 0.045$ MeV.
For the $2^{+}$ subthreshold bound state  
\begin{align}
(\gamma_{1\,0\,2\,2})^{2}=\frac{   ({\tilde \gamma}_{1\,0\,2\,2})^{2}  }{1 - ({\tilde \gamma}_{1\,0\,21\,2})^{2}\,\frac{ {\rm d}S_{2}(E_{xA}) }{  {\rm d}E_{xA} }\Big|_{ E_{xA}=-\varepsilon_{xA(2)}^{(s)}}},
\label{gamma122}
\end{align}
where $\varepsilon_{xA(2)}^{(s)}=0.2449$ MeV.  The observable reduced widths $({\tilde \gamma}_{1\,0\,1\,1})^{2}$ and $({\tilde \gamma}_{1\,0\,2\,2})^{2}$ are expressed in terms of the corresponding ANCs of the subthreshold bound states by Eq. (\ref{ANCredwidth1}). 
For the ANCs of the $1^{-}$ and $2^{+}$ subthreshold states we adopted $[C_{(\alpha\,{}^{12}{\rm C})1}^{(s)}]^{2}=4.39 \times 10^{28}$ fm${}^{-1}$ and 
$[C_{(\alpha\,{}^{12}{\rm C})2}^{(s)}]^{2}=1.48 \times 10^{10}$ fm${}^{-1}$ \cite{avila}, correspondingly.
In all the calculations, following \cite{brune1999}, we use the channel radius $R_{\alpha\,{}^{12}{\rm C}}= 6.5$ fm. 
 
The formal reduced width of the resonance $1^{-}$ 
\begin{align}
(\gamma_{2\,0\,1\,1})^{2}=\frac{({\tilde \gamma}_{2\,0\,1\,1})^{2}  }{1 - ({\tilde \gamma}_{2\,0\,1\,1})^{2}\,\frac{ {\rm d}S_{1}(E_{xA}) }{  {\rm d}E_{xA} }\Big|_{ E_{xA}=-\varepsilon_{xA(1)}^{(s)}    }}.
\label{gamma211}
\end{align}
It is important to discuss why the energy derivative $\frac{ {\rm d}S_{1}(E_{xA}) }{  {\rm d}E_{xA} }$
for the resonance $1^{-},\,E_{\alpha,\,{}^{12}{\rm C}}=2.423$ MeV is taken at $E_{xA}=-\varepsilon_{xA(1)}^{(s)}$. When we deal with a few interfering levels within the $R$-matrix approach it is convenient to adopt the boundary condition at the energy corresponding to the energy of one of the levels. Then the boundary condition for the interfering levels is taken at the same energy level for all the levels. In the case under consideration for the $E1$ transition we take into account three levels and select the boundary condition at the energy of the first level, which is the $1^{-}$ subthreshold bound state, that is, $E_{1}=-\varepsilon_{xA(1)}^{(s)}$. For the $E2$ transition we take into account two levels and select the boundary condition at the energy of the $2^{+}$ subthreshold bound state $E_{2}=-0.245$ MeV.  

The observable resonance reduced width $ ({\tilde \gamma}_{2\,0\,1\,1})^{2} $ is related to the observable resonance width as
\begin{align}
{\tilde \Gamma}_{2\,011}=2\,P_{1}(E_{R},\,R_{\alpha\,{}^{12}{\rm C}})\,({\tilde \gamma}_{2\,011})^{2}.
\label{Gammaredwidth2}
\end{align}
We adopt ${\tilde \Gamma}_{2\,011}=0.48$ MeV \cite{tiley}. 

Now we discuss the radiative width amplitudes.
The formal radiative widths are given by equations
\begin{align}
\gamma_{(\gamma)\,1\,0\,1}^{1} = {\tilde \gamma}_{(\gamma)\,1\,0\,1}^{1}\,
\sqrt{1+(\gamma_{1\,0\,1\,1})^{2}\,\frac{ {\rm d}S_{1}(E_{xA}) }{  {\rm d}E_{xA} }\Big|_{ E_{xA}=-\varepsilon_{xA(1)}^{(s)}} }\,, \nonumber\\
\gamma_{(\gamma)\,2\,0\,1}^{1} = {\tilde \gamma}_{(\gamma)\,2\,0\,1}^{1}\,
\sqrt{1+(\gamma_{2\,0\,1\,1})^{2}\,\frac{ {\rm d}S_{1}(E_{xA}) }{  {\rm d}E_{xA} }\Big|_{ E_{xA}=-\varepsilon_{xA(1)}^{(s)}} }\,\,,    \nonumber\\    
\gamma_{(\gamma)\,1\,0\,2}^{2} = {\tilde \gamma}_{(\gamma)\,1\,0\,2}^{2}\,
\sqrt{1+(\gamma_{1\,0\,2\,2})^{2}\,\frac{ {\rm d}S_{2}(E_{xA}) }{  {\rm d}E_{xA} }\Big|_{ E_{xA}=-\varepsilon_{xA(2)}^{(s)}} }\,.
\label{radwidths1}
\end{align}

\begin{figure}
[tbp] 
  \includegraphics[width=3.5in,height=2.5in,keepaspectratio]{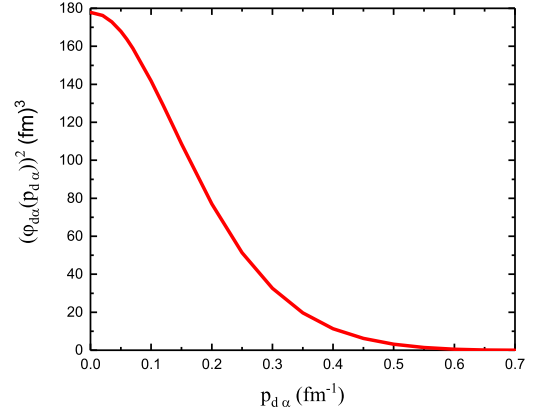}
  \caption{Square of the $d-\alpha$ bound state wave function in the momentum space.}  \label{fig_6Libstwfpspace}
\end{figure}

Another important point to discuss is the kinematics of the indirect reaction. The triple differential cross section is proportional to $\varphi_{d\,\alpha}^{2}(p_{d\,\alpha})$, which is shown in Fig. \ref{fig_6Libstwfpspace}. The maximum of $\varphi_{d\,\alpha}^{2}(p_{d\,\alpha})$ at $p_{d\,\alpha}=0$ (QF kinematics) also provides the maximum of the triple differential cross section. $p_{d\,\alpha}$ is the $d-\alpha$ relative momentum  in the three-ray vertex ${}^{6}{\rm Li} \to d + \alpha$ of the diagram in Fig. \ref{fig:fig_polediagram1}.
To calculate the Fourier transform of the $\,{}^{6}{\rm Li}=(d\,\alpha)$ bound-state wave function we use the Woods-Saxon potential with the depth $\,V_{0}= 60.0$ MeV and geometry $r_{0}=r_{C}=1.25$ fm and $a=0.65$ fm. This potential provides the $d-\alpha$ bound state with the binding energy $\,\varepsilon_{d\,\alpha}= 1.474$ MeV and the square of the ANC for the virtual decay
$\,{}^{6}{\rm Li} \to d + \alpha$  $\,[C_{(d\,\alpha)0}]^{2}= 7.28$ fm$^{-1}$. The bound-state wave number of the $\,d\,\alpha$ bound state $\kappa_{d\,\alpha}= 0.31$ fm$^{-1}$.

Usually the indirect experiments are performed at fixed incident energy of the projectiles. In the case under consideration the projectile is ${}^{6}{\rm Li}$ or ${}^{12}{\rm C}$ (in the inverse kinematics). To cover the $E_{\alpha\,{}^{12}{\rm C}}$ energy interval $\sim 2$ MeV at fixed relative kinetic energy $E_{^{6}{\rm Li}\,{}^{12}{\rm C}}$, 
one needs to change ${\rm {\bf k}}_{d}$, that is 
$p_{d\,\alpha} = |{\rm {\bf k}}_{d} - \frac{m_{d}}{m_{{}^{6}{\rm Li}}}{\rm {\bf k}}_{{}^{6}{\rm Li}}| $. For simplicity, we assume that ${\rm {\bf k}}_{d} || {\rm {\bf k}}_{{}^{6}{\rm Li}}$. 

Owing to the energy conservation by changing $k_{d}$ we can vary $E_{\alpha\,{}^{12}{\rm C}}$ but simultaneously we change the $d-\alpha$ relative momentum $p_{d\,\alpha}$. The triple
differential cross section given by Eq. (\ref{tripplediffcrsect11}) is proportional to the $d-\alpha$ bound-state wave function in the momentum space $\varphi_{d\,\alpha}^{2}(p_{d\,\alpha})$, which 
decreases with increase of $p_{d\,\alpha}$, see Fig. \ref{fig_6Libstwfpspace}.
 To avoid significant decrease of the triple differential cross section when covering the $E_{\alpha\,{}^{12}{\rm C}}$ energy interval $\approx 2$ MeV it is better to take a lower $E_{^{6}{\rm Li}\,{}^{12}{\rm C}}$ 
but not too close to the Coulomb barrier in the initial channel of the indirect reaction (\ref{THMreaction1}). Taking into account that this Coulomb barrier is $\approx 5$ MeV we consider as an example the relative kinetic energy $E_{^{6}{\rm Li}\,{}^{12}{\rm C}}=7$ MeV. 
In this case for $E_{\alpha\,{}^{12}{\rm C}}=2.28$ MeV, which is close to the resonance energy of the $1^{-}$ resonance, $p_{d\,\alpha}= 0.141$ fm${}^{-1}$ while at $E_{\alpha\,{}^{12}{\rm C}}=0.3$ MeV  $p_{d\,\alpha}= 0.281$ fm${}^{-1}$. Hence, when covering the $E_{\alpha\,{}^{12}{\rm C}}$ energy interval from the energy  $E_{\alpha\,{}^{12}{\rm C}}=2.28$ MeV to the most effective astrophysical energy for the process 
${}^{12}{\rm C}(\alpha,\,\gamma){}^{16}{\rm 16}$  the square of the Fourier transform $\varphi_{d\,\alpha}^{2}(p_{d\,\alpha})$ drops by a factor of $2.97$. Note that the drop of $\varphi_{d\,\alpha}^{2}(p_{d\,\alpha})$, when moving from $E_{\alpha\,{}^{12}{\rm C}}=2.28$ MeV to $0.9$ MeV, is $2.1$. $\,\varphi_{d\,\alpha}^{2}(p_{d\,\alpha})$ appears because we consider the indirect three-body reaction. There is another energy-dependent factor ${\tilde M}_{L}$, which is also result of the consideration of the three-body indirect reaction. This factor will be considered below.

Our main goal is to calculate the photon's angular distributions at different $E_{\alpha-{}^{12}{\rm C}}$
energies. It can allow us to compare the indirect cross sections at higher energies $E_{\alpha-{}^{12}{\rm C}} = 2.28$ and $2.1$ MeV and the most effective astrophysical energy $E_{\alpha-{}^{12}{\rm C}} = 0.3$ MeV. Because the indirect triple differential cross section does not contain the penetrability factor in the channel $\alpha-{}^{12}{\rm C}$ of the binary sub-reaction (\ref{reaction1}), the indirect method allows one to measure the triple differential cross section at $E_{\alpha-{}^{12}{\rm C}} = 0.3$ MeV what is impossible by any direct method.\\
1.  By comparing the triple differential cross sections at higher energies and at $0.3$ MeV we can determine how much the indirect cross section will drop when we reach $E_{\alpha-{}^{12}{\rm C}} = 0.3$ MeV. It will help to understand whether it is feasible to measure the triple differential cross section  at such a low energy.\\
2. The second goal is to determine whether the interference of the $1^{-}$ subthreshold resonance and $1^{-}$ resonance at $2.423$ MeV is constructive or distractive because the pattern of this interference may affect the photon's angular distribution.\\
3. The third goal is to compare the relative contribution of the $E1$ and $E2$ transitions.

\subsection{Astrophysical factors for ${}^{12}{\rm C}(\alpha,\,\gamma){}^{16}{\rm O}$}

First, to determine the parameters, which we use to calculate the triple differential cross sections, we fit the experimental  astrophysical factors $S_{E1}$ for the $E1$ transition and 
$S(E2)$ for the $E2$ transition for the ${}^{12}{\rm C}(\alpha,\,\gamma){}^{16}{\rm 16}$ reaction from \cite{redder}. We do not pursue a perfect fit and mostly are interested in fitting energies below the $1^{-}$ resonance at $2.423$ MeV,  and at low energies $E_{\alpha-{}^{12}{\rm C}} \leq 1$ MeV. To get an acceptable fit for the $E1$ transition we needed  to include three levels, two physical states, subthreshold $1^{-}$ state and the $1^{-}$ resonance, and one background state. For the $E2$ transition it was enough to include only two physical states, $2^{+}$ subthreshold resonance and $2^{+}$ resonance at $2.683$ MeV. We repeat that we do not pursue the perfect fit of the experimental $S$ factors. Our goal is to demonstrate the pattern of the triple differential cross section using reasonable parameters. More elaborated fit can be done when indirect data will be available.
 Note that in our fit we kept fixed only the parameters of the subthreshold resonances $\,1^{-}\,$ and $\,2^{+}\,$ while the parameters of the higher lying resonances $\,1^{-}\,$ and $\,2^{+}\,$ were varying.
The fixed parameters are shown in Table \ref{table_parameters} in parentheses. In this table is shown
the set of the parameters used to fit the astrophysical factors $\,S_{E1}\,$ and $\,S_{E2}\,$ and to calculate the triple differential cross section. 
\tabcolsep=3pt
\begin{table}
\begin{center}
\caption{Parameters used in calculations of the astrophysical factors of the ${}^{12}{\rm C}(\alpha,\,\gamma){}^{16}{\rm O}$ radiative capture and the photon's angular distributions from the indirect ${}^{12}{\rm C}({}^{6}{\rm Li}, d\,\gamma){}^{16}{\rm O}$ reaction.}
\vspace{0.4cm}
\begin{tabular}[t]{c c c}
\hline \hline \\
 &  $L=1$ &  $L=2$  \\[2mm]
\hline\\[1mm]
$E_{1}$ [MeV] & $(-0.45)$ &  $(-0.245)$   \\[1ex]
 $\gamma_{1\,0\,L\,L}$ [MeV$^{1/2}$]\,\,\, & $(0.0867)$ & $(0.1500)$  \\[1ex] 
  $\gamma_{(\gamma)\,1\,0\,L\,L}^{L}$ [MeV${}^{1/2}$fm${}^{L+1/2}$] & $(0.0241)$ & $(0.9415)$  \\[1ex]
    $E_{2}$ [MeV] & $3.0$ & $2.8$  \\[1ex]
   $\gamma_{2\,0\,L\,L}$ [MeV$^{1/2}$] & $0.3254$ & $0.75$  \\[1ex]
$\gamma_{(\gamma)\,2\,0\,L\,L}^{L}$ [MeV${}^{1/2}$fm${}^{L+1/2}$]  & $-0.00963$ & $-0.09257$  \\[1ex]
$E_{3}$ [MeV] & $33.8$ &   \\ [1ex]  
$\gamma_{3\,0\,L\,L}$ [MeV$^{1/2}$]  & $1.1$ &  \\[1ex]
$\gamma_{(\gamma)\,3\,0\,L\,L}^{L}$ [MeV${}^{1/2}$fm${}^{L+1/2}$] & $-0.00239$ &  \\[2mm]
\hline \hline \\
\label{table_parameters}
\end{tabular}
\end{center}
\end{table}
$E_{n}$ is the energy of the $n$-th level. 
Note that in the $R$-matrix approach, which includes a few interfering levels, it is convenient to choose one of the energy levels coinciding with the location of the observable physical state. 

In this paper we adopted $E_{1}=-\varepsilon_{\alpha\,{}^{12}{\rm C}(1)}^{(s)}=-0.045$ MeV for $L=1$ and $E_{2}=-\varepsilon_{\alpha\,{}^{12}{\rm C}(2)}^{(s)}= -0.245$ MeV for $L=2$ transitions. 
Then the boundary condition for the second and third levels of the $E1$ transition are taken at $E_{\alpha\,{}^{12}{\rm C}}=-0.045$ MeV and for $L=2$ the boundary condition is taken at $E_{\alpha\,{}^{12}{\rm C}}= -0.245$ MeV. Moreover, because in our choice the locations of the subthreshold bound states for $L=1$ and $L=2$ are fixed, the energies of other levels are  fitting parameters and deviate from the real resonance energies. For example, the $1^{-}$ resonance at $2.423$ MeV in the fit is shifted to $E_{\alpha\,{}^{12}{\rm C}}= 3.0
$ MeV and the $2^{+}$ resonance at $2.683$ MeV is shifted to $2.8$ MeV. Hence, the statement that we take into account the radiative capture through the wing of the subthreshold $1^{-}$ resonance at $E_{\alpha\,{}^{12}{\rm C}}=-0.045$ MeV and the $1^{-}$ resonance at $E_{\alpha\,{}^{12}{\rm C}}= 2.423$ MeV does not contradict to the fact that in the fit the resonance at $2.423$ MeV is shifted to $3.0$ MeV. To fit the $E1$ transition we needed to add the bacground state at $33.8$ MeV with parameters given in Table \ref{table_parameters}. 

In this table the given parameters provide the constructive interference of the subthreshold $1^{-}$ resonance and resonance at $2.423$ MeV at low energies. Changing the sign of 
$\gamma_{(\gamma)\,2\,0\,1\,L}^{1}=-0.00963$ MeV${}^{1/2}$fm${}^{3/2}$ to positive  provides the destructive interference between the first two $1^{-}$ levels. In what follows by the $E1$ 
constructive (destructive) interference we mean the constructive (desctructive) interference between the first two $1^{-}$ levels.  

\begin{figure}
[tbp] 
  \includegraphics[width=6.0in,height=2.5in,keepaspectratio]{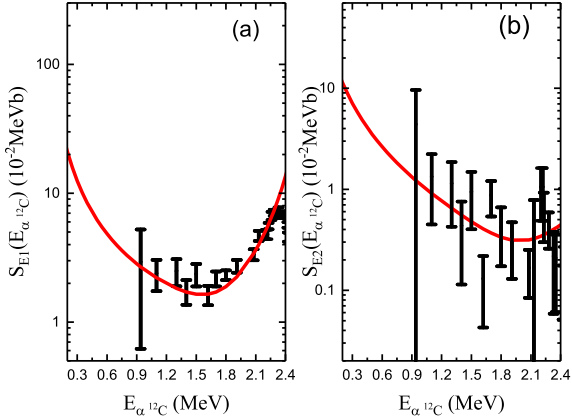}
  \caption{Low-energy astrophysical $S_{E1}(E_{\alpha\,{}^{12}{\rm C}})$ and $S_{E2}(E_{\alpha\,{}^{12}{\rm C}})$ factors for $E1$ and $E2$ transitions for the  ${}^{12}{\rm C}(\alpha,\,\gamma){}^{16}{\rm O}$ radiative capture. Black dotes are astrophysical factors from \cite{redder}, solid red line is present paper fit.
Panel (a): $S_{E1}(E_{\alpha\,{}^{12}{\rm C}})$ astrophysical factor; panel (b): $S_{E2}(E_{\alpha\,{}^{12}{\rm C}})$ astrophysical factor.}  
\label{fig_Sfactors}
\end{figure}

In Fig. \ref{fig_Sfactors} the calculated $S_{E1}$ and $S_{E2}$ astrophysical factors for the $E1$ and $E2$ transitions, correspondingly, are compared with the experimental ones from \cite{redder}. 
Our fitted astrophysical factors are: $S_{E1}(0.3\,{\rm MeV})= 124.6$ keVb for the $E1$ transition and $S_{E2}(0.3\,{\rm MeV})= 71.1$ keVb for the $E2$ transition. Evidently that our 
value for the $E1$ transition is higher than the contemporary accepted value of $80$ keVb for constructive interference but the value for the $E2$ transition is close to the low value $60$ keVb \cite{gai2015}. But, as we have underscored, our values should not be taken very seriously.
In the absence of indirect data we use the parameters obtained from fitting the data from \cite{redder} to generate the photon's angular distributions to make some qualitative predictions.
We also show how the photon's angular distributions are affected by lowering $S_{E1}(0.3\,{\rm MeV})$.
 
\subsection{Photon's angular distributions}

\begin{figure}
[tbp]   
   \includegraphics[width=6.0in,height=2.5in,keepaspectratio]{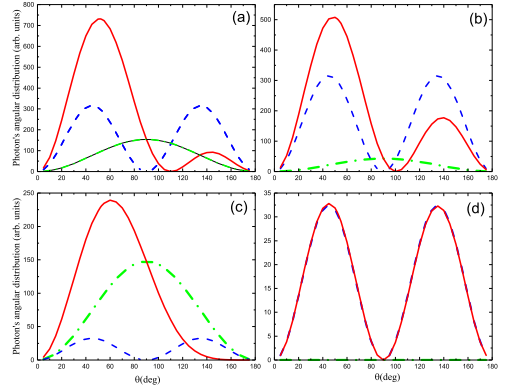}
  \caption{Angular distribution of the photons emitted from the reaction ${}^{12}{\rm C}({}^{6}{\rm Li},d\,\gamma){}^{16}{\rm O}$ proceeding through the wings of two subthreshold resonances $1^{-},\,E_{\alpha\,{}^{12}{\rm C}} = -0.045$ MeV and $2^{+},\,E_{\alpha\,{}^{12}{\rm C}}= -0.245$ MeV, and the resonances at $E_{\alpha\,{}^{12}{\rm C}} >0$. The green dashed-dotted line is the angular distribution for the electric dipole transition $E1$, the blue dashed line is the angular distribution generated by the electric quadrupole $E2$ transition, and the red solid line is the total angular distribution resulted from the interference of the $E1$ and $E2$ radiative captures.
Panel (a):  $\,E_{\alpha\,{}^{12}{\rm C}}= 0.3$ MeV, constructive interference of the $\,E1$ transitions through the wing of $\,1^{-},\,E_{\alpha\,{}^{12}{\rm C}} = -0.045$ MeV and the resonance $\,1^{-},E_{R}= 2.423$ MeV; panel (b): $\,E_{\alpha\,{}^{12}{\rm C}}= 0.3$ MeV, destructive interference of the $E1$ transitions through the wing of $\,1^{-},\,E_{\alpha\,{}^{12}{\rm C}} = -0.045$ MeV and the resonance $\,1^{-},E_{R}= 2.423$ MeV;  panel (c): the same as panel (a) for $\,E_{\alpha\,{}^{12}{\rm C}}= 0.9$ MeV; panel (d): the same as panel (b) for $\,E_{\alpha\,{}^{12}{\rm C}}= 0.9$ MeV.}  \label{fig_Angdistr1}
\end{figure}

\begin{figure}
[tbp] 
 \includegraphics[width=6.0in,height=2.5in,keepaspectratio]{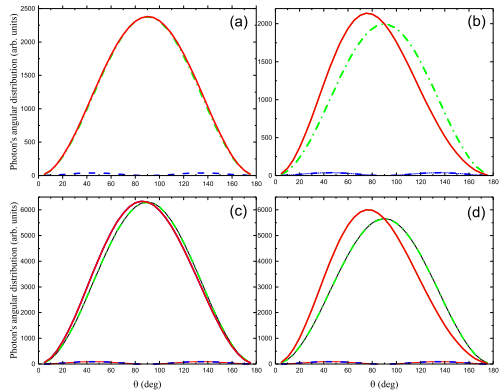}
 \caption{Angular distribution of the photons emitted from the reaction ${}^{12}{\rm C}({}^{6}{\rm Li},d\,\gamma){}^{16}{\rm O}$ proceeding through the wings of the two subthreshold resonances, $1^{-},\,E_{\alpha\,{}^{12}{\rm C}} = -0.045$ MeV and $2^{+},\,E_{\alpha\,{}^{12}{\rm C}}= -0.245$ MeV,  and the resonances at $E_{\alpha\,{}^{12}{\rm C}} >0$. Notaions of the lines are the same as in Fig. \ref{fig_Angdistr1}. 
Panel (a): the same as panel (a) in Fig. \ref{fig_Angdistr1} for $E_{\alpha\,{}^{12}{\rm C}}= 2.1$ MeV; panel (b): the same as panel (b) in Fog. \ref{fig_Angdistr1} for $E_{\alpha\,{}^{12}{\rm C}}= 2.1$ MeV;
panel (c) the same as panel (c) in Fig. \ref{fig_Angdistr1} for $E_{\alpha\,{}^{12}{\rm C}}= 2.28$ MeV.}\label{fig_Angdistr2}
\end{figure}

\begin{figure}
[tbp] 
 \includegraphics[width=6.0in,height=2.5in,keepaspectratio]{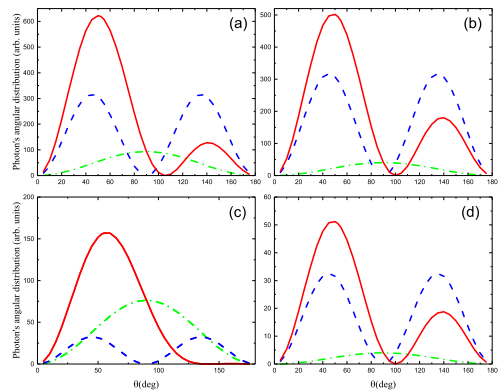}
 \caption{The same as in Fig.  \ref{fig_Angdistr1} but the calculations are done with three modified $R$-matrix parameters generating lower $S_{E1}(0.3)$ MeV astrophysical factor. }  \label{fig_Angdistr3}
\end{figure}

\begin{figure}
[tbp] 
 \includegraphics[width=6.0in,height=2.5in,keepaspectratio]{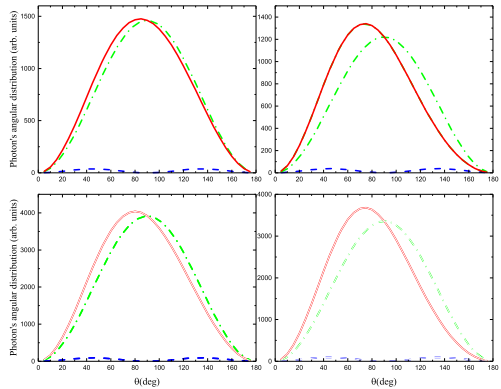}
 \caption{The same as in Fig.  \ref{fig_Angdistr2} but calculations are done with three modified $R$-matrix parameters generating lower $S_{E1}(0.3)$ MeV astrophysical factor.}  \label{fig_Angdistr4}
\end{figure}

In Figs. \ref{fig_Angdistr1}, \ref{fig_Angdistr2}, \ref{fig_Angdistr3} and \ref{fig_Angdistr4} the photon's angular distributions are shown at four
different $\,E_{\alpha\,{}^{12}{\rm C}}\,$ energies: $\,0.3,\,0.9,2.1$ and $\,2.28$ MeV. 
We don't show the angular distributions at middle energies, for example, at $\,1.5$ MeV, because it turns out that the result at this energy is quite sensitive to the adopted channel radius. The calculations are performed at $\,E_{{}^{6}{\rm Li}\,{}^{12}{\rm C}}=7$ MeV ($9.33$ MeV in the Lab. system with $\,{}^{6}{\rm Li}\,$ projectile), which is higher than the Coulomb barrier $\,V_{CB} \approx 5\,$ MeV in the entry channel $\,{}^{6}{\rm Li} + {}^{12}{\rm C}\,$ of the indirect reaction. 

Figures \ref{fig_Angdistr1} and \ref{fig_Angdistr2} are very instructive. 
First, we note that the $E1$ angular distributions of the photons at all energies are peaked at $90^{\circ}$ while the $E2$ angular distributions are double-humped and peaked at $45^{\circ}$ and $135^{\circ}$. However, the interference of the $E1$ and $E2$ transitions leads to different total angular distributions.
The angular distributions at $0.3$ MeV are quite similar for the $E1$ transitions with constructive and destructive interferences, panels $(a)$ and $(b)$ in Fig. \ref{fig_Angdistr1}, with pronounced peaks at $52^{\circ}$ and $50^{\circ}$, correspondingly. The character of the total angular distribution at $0.3$ MeV depends on the relative weight of the $E1$ and $E2$ transitions.

In the sense of the distinguishing between the constructive and destructive $E1$ transitions,  the photon's angular distributions at $0.9$ MeV, panels $(c)$ and $(d)$, are the most instructing: the patterns of the photon's angular distributions are quite different for the constructive and destructive $E1$ transitions what allows one to distinguish between two types of the $E1$ interferences. However, the cross sections for the destructive $E1$ interference is too small compared to the cross section at $0.3$ MeV.

Now we proceed to the angular distributions at higher energies shown in Fig. \ref{fig_Angdistr2}. At higher energies the $E1$ transition dominates and we see profound $E1$ type angular distributions both for the $E1$ constructive and destructive interferences of the two first $1^{-}$ levels. Hence, the angular distributions at higher energies cannot distinguish between constructive and destructive $E1$ interferences.

Comparing the relative values of the triple differential cross sections 
of Fig. \ref{fig_Angdistr2}, panel (c) and Fig. \ref{fig_Angdistr1}, panel (a) we can make, presumably, the most important conclusion: 
the triple differential cross section near the $1^{-}$ resonance at $2.28$ MeV exceeds the one at $0.3$ MeV by approximately an order of magnitude. We remind to the reader that in the case of the direct measurements when moving from $2.28$ MeV to $0.3$ MeV the cross section drops by a factor of  $10^{9}$. Our estimation detailed in the next section shows that measurements of the indirect triple differential cross section at $0.3$ MeV are feasible. Thus, for the first time, we provide a possibility  to measure the ${}^{12}{\rm C}(\alpha,\,\gamma){}^{16}{\rm O}$ right at the most effective astrophysical energy $0.3$ MeV.  

In Figs  \ref{fig_Angdistr1} and \ref{fig_Angdistr2} we have used the $R$-matrix parameters, which 
provide a higher $S_{E1}(0.3\,{\rm MeV})= 124.6$ keVb for the constructive $E1$ transition than the contemporary accepted  $\sim 80$ keVb \cite{gai2015}. To check how the photon's angular distributions are affected by a lower $E1$ astrophysical we changed three $R$-matrix parameters
in Table \ref{table_parameters}:  $E_{2} =3.1$ MeV,  $\,\gamma_{(\gamma)\,2\,0\,1\,1}^{1} = -0.006132$ MeV${}^{1/2}$fm${}^{3/2}$ and  $\,\gamma_{3\,0\,1\,1}=1.4$ MeV$^{1/2}$.
With these parameters we get $S_{E1}(0.3\,{\rm MeV})= 75.8$ keVb and $S_{E1}(0.9\,{\rm MeV}) =14.7$ keVb. We use the modified parameters to calculate the photon's angular distributions again 
at $E_{\alpha\,{}^{12}{\rm C}}=0.3,\,0.9,\,2.1$ and $2.28$ MeV, see Figs \ref{fig_Angdistr3} and \ref{fig_Angdistr4}. Thus we repeated calculations similar to the ones shown in Figs. \ref{fig_Angdistr1} and \ref{fig_Angdistr2} but with three modified parameters leading to smaller $S_{E1}$.


We find that decrease of the $S_{E1}$ does not change the angular distribution except for the panel (d) in \ref{fig_Angdistr3}, which is quite different than the panel (d) in Fig \ref{fig_Angdistr1}  but the absolute values of the cross sections in these panels are quite small. The main effect of decreasing of the $S_{E1}$factor is decrease of the triple differential cross section at higher energies where $E1$ significantly dominates over $E2$. As the result,
the ratio of the triple differential cross sections at $2.28$ MeV and $0.3$ MeV is only $6.5$, that is, the relative weight of the triple differential cross section at $0.3$ MeV increases what makes more plausible the chances to measure the triple differential cross section at $0.3$ MeV for lower $S_{E1}$. 

In \cite{gai2015} it was underscored that contemporary experimental data do not exclude 
very low $S_{E1}(0.3\,{\rm MeV})=10$ keVb and high $S_{E2}(0.3\,{\rm MeV})=154$ keVb. We did not exploit here all the possibilities for the astrophysical factors but, evidently that this marginal values can change significantly the photon's angular distributions. Indirect measurements can finally resolve ambiguities in the low-energy astrophysical factors.

\section{Feasibility of the proposed approach}

Reliable estimates for the $^{12}{\rm C}({}^{6}{\rm Li},d){}^{16}{\rm O}$ reaction cross section at 10-11 MeV energy of $^6$Li beam populating the $1^-$ state at 9.585 MeV can be made. Using FRESCO  reaction code \cite{fresco} and the same set of potentials as in \cite{avila}, DWBA calculations predict cross section of this reaction on the order of 10 mb/sr for forward angles (0-30$^\circ$ in c.m.). The  $\gamma$ branching of this state to the $^{16}$O ground state is 5$\times$10$^{-8}$ \cite{tiley}. This sets the absolute scale for the cross sections to be measured at close to 1 nb. This is a very challenging but achievable target for a dedicated experimental setup. One possibility is to couple high efficiency array for high energy $\gamma$-rays (such as BaF) with large area position sensitive Si array to detect deuterons. Another possibly is to use inverse kinematics ($^{12}{\rm C}$ beam on $^6$Li target) and detect $^{16}$O recoils in the spectrometer while still measuring deuterons at back angles in coincidence with high energy $\gamma$-rays. We estimate that event rates as high as 10$^3$ per day can be achieved with high intensity beams (on the order of 1 particle $\mu$A) while keeping energy resolution within 100 keV. Slow variation of the triple differential cross section with energy (by one order of magnitude) makes it possible to achieve satisfactory statistics even at $E_{\alpha\,^{12}{\rm C}} = 0.3$ MeV within reasonable time frame (one-two weeks of beam time).

\section{Summary}

In this paper we suggested and developed the formalism of resonant indirect radiative capture reactions. The derived expressions for the triple and double differential cross sections   
can be used for the analysis of the indirect radiative capture reactions. The developed formalism can be used when indirect reactions proceed through a few subthreshold bound states and resonances. In this case the statistical theory cannot be applied and the intermediate subthreshold bound states and resonances should be taken into account explicitly. 

The idea of the indirect method is to use the indirect reaction $A(a, s\,\gamma)F$ to obtain the information about the radiative capture reaction $A(x,\,\gamma)F$, where $a=(s\,x)$ and $F=(x\,A)$. The main advantage of using the indirect reactions is the absence of the Coulomb-centrifugal penetrability factor in the entry channel $x+A$ of the binary sub-reaction $A(x,\,\gamma)F$, which suppresses the low-energy cross section of this reaction and does not allow one to measure it at astrophysically relevant energies. 

Using indirect resonant radiative capture reactions one can obtain the information about important astrophysical resonant radiative capture reactions, like $(p,\,\gamma), \,\,(\alpha,\,\gamma)$ and $(n,\,\gamma)$ on stable and unstable isotopes. The indirect technique makes accessible low-lying resonances, which are close to the threshold, and even subthreshold bound states at negative energies. 

In this paper, after developing the general formalism, we have demonstated the application of the indirect method for the indirect reaction ${}^{12}{\rm C}({}^{6}{\rm Li},d\,\gamma){}^{16}{\rm O}$ proceeding through $1^{-}$ and $2^{+}$ subthreshold bound states and resonances to obtain the information about the ${}^{12}{\rm C}(\alpha,\,\gamma){}^{16}{\rm O}$ radiative capture.

The indirect method requires measurement of the triple differential cross section in the coincidence experiment, in which the photon's angular distribution is measured at given
energy and scattering angle of the deuteron. This photon's angular distribution is the photon-deuteron angular correlation.

We show that the ratio of the triple differential cross section  at energy $E_{\alpha\,{}^{12}{\rm C}}=2.28$ MeV, which is close to the $1^{-}$ resonance at $2.423$ MeV, to the one at  $E_{\alpha\,{}^{12}{\rm C}}=0.3$ MeV is about an order of magnitude. Such a small drop of the triple differential cross section when one reaches the most effective astrophysical energy 
$E_{\alpha\,{}^{12}{\rm C}}=0.3$ MeV makes it possible to obtain the information about the astrophysical factor for the ${}^{12}{\rm C}(\alpha,\,\gamma){}^{16}{\rm O}$ process. We remind that in the direct experiment the cross section of the ${}^{12}{\rm C}(\alpha,\,\gamma){}^{16}{\rm O}$ reaction drops by $\,\sim 10^{9}$ when moving from the energies close to the resonance at $\,2.423$ MeV down to $\,0.3$ MeV.
We discuss also the optimal experimental kinematics to measure the indirect reactions and, in particular, the ${}^{12}{\rm C}({}^{6}{\rm Li},\,d\,\gamma){}^{16}{\rm O}$ process.

\section{Acknowledgments}

A.M.M. and G.V.R. acknowledge support from the U.S. DOE Grants
No. DE-FG02-93ER40773. A.M.M. also acknowledges the support
by the U.S. NSF Grant No. PHY-1415656. G.V.R. also acknowledges the financial support of the Welch
Foundation (USA) (Grant No. A-1853).

\end{document}